\documentclass[fleqn,usenatbib]{mnras}

\usepackage{showyourwork}

\usepackage{newtxtext,newtxmath}
\usepackage[utf8]{inputenc}
\usepackage[T1]{fontenc}
\usepackage{graphicx}	
\usepackage{amsmath}	
\usepackage{xspace}
\usepackage{subfig}
\usepackage[nameinlink]{cleveref}
\usepackage{dsfont}
\usepackage{enumitem}
\usepackage{tikzit}

\newcommand{\tbf}[1]{\textbf{#1}}

\newcommand{\norm}[1]{\lVert #1 \rVert}
\newcommand{\mbf}[1]{\mathbf{#1}}
\newcommand{\mboxbf}[1]{\mbox{\boldmath$#1$}}
\newcommand{\mfrak}[1]{\mathfrak{#1}}
\newcommand{\mcal}[1]{\mathcal{#1}}

\newcommand{\stream}[1]{#1}
\newcommand{\nbody}{$N$-body simulation}
\newcommand{\nbodys}{$N$-body simulations}
\newcommand{\gaia}{\texttt{Gaia}}

\newcommand{\package}[1]{\texttt{#1}\xspace}

\newcommand{\python}{\package{Python}}

\newcommand{\trackstream}{\package{TrackStream}}

\newcommand{\astropypkg}{\package{astropy}}

\newcommand{\astropyapi}[2]{\href{https://docs.astropy.org/en/stable/api/astropy.#1.html}{#2}}
\newcommand{\astropyapidoc}[2]{\astropyapi{#1}{\texttt{#2}\xspace}}

\newcommand{\astropySkyOffsetFrame}{\astropyapidoc{coordinates.builtin_frames.SkyOffsetFrame}{SkyOffsetFrame}}

\newcommand{\galpy}{\package{galpy}}
\newcommand{\galpyapi}[2]{\href{https://docs.galpy.org/en/stable/#1.html}{#2}}
\newcommand{\galpyapidoc}[2]{\galpyapi{#1}{\texttt{#2}\xspace}}
\newcommand{\galpystreamdf}{\galpyapidoc{streamdf}{streamdf}}
\newcommand{\galpystreamspraydf}{\galpyapidoc{streamdf}{streamspraydf}}
\newcommand{\galpyMWPotential}{\href{https://docs.galpy.org/en/stable/reference/potential.html?highlight=MWPotential2014}{\texttt{MWPotential2014}}{}}

\newcommand{\nbodysix}{\texttt{NBODY6}}

\tikzstyle{data}=[fill={rgb,255: red,4; green,255; blue,0}, draw=none, shape=circle, inner sep=1pt]
\tikzstyle{SOM node}=[fill={rgb,255: red,137; green,137; blue,137}, draw=black, shape=circle, inner sep=0pt]
\tikzstyle{label}=[fill=white, draw=none, shape=rectangle, rotate=22]

\tikzstyle{path}=[draw=green, -, fill=none]
\tikzstyle{projection}=[-, dashed, ultra thin]
\tikzstyle{convex sweep}=[fill={rgb,255: red,155; green,155; blue,255}, draw={rgb,255: red,0; green,17; blue,255}, dashed, ultra thin, ->, fill opacity=0.5]
\tikzstyle{region 1}=[-, fill={rgb,255: red,255; green,191; blue,191}, draw=none, fill opacity=0.5]
\tikzstyle{region 2}=[-, fill={rgb,255: red,255; green,252; blue,152}, draw=none, fill opacity=0.5]
\tikzstyle{region 3}=[-, fill={rgb,255: red,207; green,255; blue,255}, draw=none, fill opacity=0.5]
\tikzstyle{direction}=[fill=none, ->]

\DeclareRobustCommand{\VAN}[3]{#2} \let\VANthebibliography\thebibliography
\def\thebibliography{\DeclareRobustCommand{\VAN}[3]{##3}\VANthebibliography}

\title{On the Fast Track: Rapid construction of stellar stream paths}

\author[N. Starkman et al.]{ Nathaniel Starkman$^{1}$\thanks{E-mail:
n.starkman@mail.utoronto.ca}, Jo Bovy$^{1}$, Jeremy J. Webb$^{1}$, Daniela
Calvetti \& Erkki Somersalo,$^{2}$ \\
$^{1}$David A. Dunlap Department of Astronomy and Astrophysics, University of
Toronto, 50 St. George Street, Toronto, Ontario, M5S 3H4, Canada\\
$^{2}$Department of Mathematics, Case Western Reserve University, Cleveland,
Ohio, USA\\
}

\date{Accepted XXX. Received YYY; in original form ZZZ}

\pubyear{2022}

\begin{document}
\pagerange{\pageref{firstpage}--\pageref{lastpage}}
\maketitle

\begin{abstract}
  Stellar streams are sensitive probes of the Galactic potential. The likelihood
  of a stream model given stream data is often assessed using simulations.
  However, comparing to simulations is challenging when even the stream paths
  can be hard to quantify. Here we present a novel application of
  Self-Organizing Maps and first-order Kalman Filters to reconstruct a stream's
  path, propagating measurement errors and data sparsity into the stream path
  uncertainty. The technique is Galactic-model independent, non-parametric, and
  works on phase-wrapped streams. With this technique, we can uniformly analyze
  and compare data with simulations, enabling both comparison of simulation
  techniques and ensemble analysis with stream tracks of many stellar streams.
  Our method is implemented in the public \python package \trackstream{},
  available at \url{https://github.com/nstarman/trackstream}.
\end{abstract}

\begin{keywords}
  Astrophysics - Astrophysics of Galaxies -- Galaxy: structure -- Galaxy:
  kinematics and dynamics -- methods: data analysis -- methods: statistical
\end{keywords}



\label{firstpage}

\section{Introduction} \label{sec:intro}

  A stellar stream is the association of stars tidally stripped from a common
  progenitor system that orbits in an underlying galactic gravitational field
  \citep{Johnston1998, HelmiWhite1999}. Stream progenitors are gravitationally
  bound systems, such as satellite dwarf galaxies or star clusters
  \citep{Odenkirchen2001, Majewski2003}. In the absence of external interactions
  these systems are stable, though still capable of mass loss driven by internal
  processes \citep{Hills1975, HeggieHut2003}, e.g. relaxation and binary
  interactions. Considering external interactions, these progenitor systems may
  be disrupted: completely, as in the merger events that formed the Galactic
  stellar halo \citep{Lynden-Bell1967, Searle1978}; or partially, such as tidal
  disruption forming a stellar stream.

  In the approximation of a smooth galactic gravitational field, the
  progenitor's extent is limited to the tidal radius. Through processes internal
  and external, progenitor stars can achieve sufficient energy to bring them
  beyond the tidal boundary, thus ``escaping'' into the galactic field. Escape
  primarily occurs at both an inner and outer saddle point of the combined
  potential -- e.g. the L2 and L3 Lagrange points for a spherical galactic field
  \citep{Ross1997, Fukushige2000, BinneyTremaine2008}. Generally, the initial
  phase space position of the escaped star is very similar to that of the
  progenitor: the configuration position is the tidal boundary and the star has
  a low peculiar velocity. Since both the escaped star and its progenitor orbit
  in the same underlying galactic potential and have similar phase space
  positions their orbits are similar \citep{BinneyTremaine2008}. Given
  sufficient time the phase space separation will increase and an escaped star
  will lead or trail the position of the progenitor, depending on the location
  from which it escaped. Therefore, a collection of tidally stripped stars will
  form extended structures, aka tails or streams, that are few to thousands of
  parsecs in length \citep{Johnston1998, HelmiWhite1999, Bovy2014}.

  To be a coherent, observable structure the peculiar velocity dispersion of
  escaped stars relative to the progenitor must be relatively low, but does vary
  from system to system. Stars in kinematically hot streams (e.g. from
  progenitors with larger kinematic dispersions) have more dissimilar orbits
  from each other and the progenitor than do stars in cold streams, with smaller
  dispersions \citep{Johnston1998, Hendel2015}. The orbit of a star depends on
  its phase space position and the gravitational potential; therefore measuring
  a star's position and velocity and observing its orbit reveals the local
  potential, which is normally dominated by the Galaxy (primarily the dark
  matter halo at large distances) \citep{Gibbons2014}. However, an orbit is
  impractical to observe on human timescales. Cold streams offer a workaround.
  Like measuring one star at many points in one orbit over an extended period,
  streams are many stars measured at different points along approximately one
  orbit at the same time. Therefore stellar streams are one of the most
  promising means to study the Galactic gravitational potential: the stellar
  components as well as the dark matter \citep{Johnston2016}.
  
  On the observational front there has been a large concerted effort in the
  field to find new stellar streams \citep[for an extensive list of known
  Galactic streams see][]{Mateu2022}, extend the detected extents of known
  streams \citep[e.g. for \stream{Palomar 5} alone:][]{Rockosi2002, Ibata2017,
  Grillmair2006, Carlberg2012, Starkman2019}, and increase the general purity of
  stream catalogs. On the simulation end, there are numerous methods to generate
  and model stellar streams, spanning many degrees of complexity. The simplest
  method is modeling the stream as tracers sampled from a single orbit, a
  technique called orbit fitting \citep[e.g.][]{Johnston1999, Malhan2019}.
  However, streams generally do not lie on a single orbit
  \citep{SandersBinney2013I} and this approximation can produce biased results.
  The distribution function method of \citet{Bovy2014} starts from the orbit of
  the progenitor, but also models the mean difference between the stream's and
  progenitor's orbits. Like orbit fitting, this method produces an analytic
  characterization of the stream's trajectory, albeit in action-angle
  coordinates that are converted to a phase-space track by linear interpolation.
  The distribution function method is a marked improvement over the orbit
  approximation, however it makes a number of simplifying assumptions and also
  struggles to model interactions with the stream that can perturb the path.
  Particle spray models \citep[e.g.][]{Fardal2015,Bonaca2014} redress some of
  these shortcomings by integrating orbits of many tracer particles released
  over time from a simulated progenitor. Models of tidal disruption and
  interactions can be explored in this framework. However, particle spray
  methods are non-analytic and are also more computationally intensive than
  their analytic counterparts. The most expensive, but also most accurate and
  detailed, stream modeling method is direct \nbody{} simulations, e.g. bespoke
  modeling of \stream{GD-1} to try to identify the progenitor \citep{Webb2019}.
  
  How observations and simulations are compared depends on the simulation
  method. When using analytic methods like distribution functions
  \citep{Bovy2014} this comparison can unsurprisingly be done analytically, by
  directly modelling the mean path, width, and other properties along the
  stream. Computing fit statistics to real data is straightforward in this
  framework. For non-analytic simulation methods, where the stream is modeled by
  a set of tracer particles, a direct comparison to data is challenging. One
  method, from \citet{Bonaca2014}, is to calculate probabilities for each tracer
  and marginalize over all model points. For a smooth likelihood function this
  method requires several times more model points than data points, and is a
  notable limitation when trying to closely match observations, e.g. with an
  N-body simulation. Thus, this method is well-suited for its intended use case,
  less so for other situations. A different and common approach to analyzing
  streams, both in simulations and from observations, is by fitting a stream
  track: a smooth path through phase space which characterizes the trajectory of
  the stream. For visualization, checking the sky footprint, and calculating
  model likelihoods the track of a stream is a multi-purpose tool. In many
  respects, what is meant by a stellar stream is its track, not its constituent
  stars.
  
  Despite its centrality to the field, there is no set definition and a
  multitude of computational methods for defining a stream track. The most
  common definition of a stream track is with a low order polynomial fit. The
  stream is projected from its initial reference frame, e.g. ICRS
  \citep{ICRS1997}, into a rotated on-sky frame, with longitude $\phi_1$ and
  latitude $\phi_2$, such that the stream extends along $\phi_1$. In this frame
  the stream may be fit by a low-order polynomial as a function of $\phi_1$
  \citep[e.g.][]{Bonaca2020}. Similar, if more sophisticated, methods swap tools
  like univariate splines \citep[e.g.][]{Erkal2017, Bonaca2018} for the
  low-order polynomials. Using a coordinate dimension, here $\phi_1$, as the
  independent variable works for many streams, as $\phi_1$ sweeps monotonically
  along the stream. For example \stream{GD-1} is straight for over 80 degrees
  \citep{Webb2019, Price-Whelan2018}. However, this is not true in general: a
  stream's path can become kinked by subhalo interactions, or curved on the
  celestial sphere in an observer's reference frame, or like the Magellanic
  stream wrap multiple times around the Milky Way \citep{Wannier1972}. When a
  stream track is not a function, methods fitting $\phi_2(\phi_1)$ fail. One
  solution is to instead fit in both $\phi_1$ and $\phi_2$ non-parametrically,
  for example using 2-D splines \citep{Erkal2017,Koposov2019, Li2021,
  Tavangar2022}. This method allows for the path, width, and numerous other
  properties to be fit simultaneously. Though implemented in 2-D, the method is
  extensible to higher dimensions and so can be used, e.g. with full 3-D
  measurements from \gaia{} or simulations. While powerful, this technique is
  slow, requiring Monte Carlo methods to optimize the number and placement of
  the spline nodes. Given its speed constraints, the method cannot be reasonably
  used as a step in a larger Monte Carlo analysis, such as to derive stream
  tracks for the generative stream models discussed previously.
  
  In order to compare simulations to data as a step in a Monte Carlo analysis we
  require a fast method to derive stream tracks. Without the opportunity to
  examine each step, the method must work when the stream track is not a
  one-to-one function, such as with phase wraps. We require the method to work
  on data with errors, constrained to be on-sky or in full 3-D, and with and
  without velocities. Furthermore, the same method should work on many different
  mock stream generation methods -- including analytic, particle sprays, and
  \nbody{} simulations -- so that results derived with one simulation technique
  can be compared to the results from another. In this paper we present a novel
  method for accomplishing these goals: rapidly constructing stellar stream
  tracks, accounting for measurement error and data sparsity. The technique is
  Galaxy-model independent, non-parametric, and works on many different mock
  stream generation methods. Our method is implemented in the public Python
  package \trackstream{}, available at
  \url{https://github.com/nstarman/trackstream}. 

  \vspace{5pt}

  The organization of this paper is as follows. In
  \autoref{sec:ordering_the_data} we present a sequence of methods for ordering
  the stars in a stellar stream. This order will allow us to treat the data
  similarly to a time-ordered data set, and therefore apply time-series
  techniques to reconstruct the stream track. In \autoref{sec:kalman_filter} we
  present one such application -- the Kalman Filter. In \autoref{sec:results} we
  apply the stream track pipeline to a variety of data. We conclude with
  discussion in \autoref{sec:discussion}.



\section{Methods: Ordering the Data} \label{sec:ordering_the_data}

  The first step in the stream path fitting procedure is ordering the data. With
  simulated data, there are numerous options for ordering, such as the ejection
  time from the progenitor \citep{Gibbons2014} or in action-angle coordinates
  \citep{Bovy2014}. However real observations rarely, if ever, permit these
  orderings. Instead we order using only configuration and kinematic phase-space
  information.

  Consider the toy model of a static, spherical potential (for both the Galaxy
  and progenitor) where all test masses (stars) are ejected with the same
  velocity. As the progenitor orbits the Galactic COM, its orbital path
  describes an elliptical path in an orbital plane. The progenitor's tidal tails
  (stream) will likewise lie near the progenitor's orbital plane. Observing from
  the Galactic center, an orbital plane lies on a great circle in the observer's
  `sky' and it is possible to re-orient the longitude ($\ell$) and latitude
  ($b$) to a stream-oriented coordinate system ($\phi_1, \phi_2$) with the
  progenitor at the origin and the stream along $\phi_1 (\phi_2 \approx 0)$.
  However, our observations are not from the Galactic center, but approximately
  8 kpc distant \citep[for recent measurements, see][]{GRAVITY2018, Leung2022}.
  For an arbitrary stream lying (approximately) along a galactocentric ellipse
  there is no guaranteed rotation to ``linearize'' the stream in the observer
  coordinates so that the stream lies along $\phi_2 \approx 0$. Thankfully, many
  streams are sufficiently distant from both the Galaxy and observer or oriented
  such that a rotation a) exists, or b) applies to significant lengths of the
  stream before deviations become problematic. When modeling streams and fitting
  a track in the same representation as the data, for most streams transforming
  to the stream-oriented coordinate system is a worthwhile first step to
  ordering the data and is discussed in
  \autoref{sub:great_circle_coordinate_frame}.
  
  The second step takes advantage of the low-dimensional structure of streams.
  Spatially, streams are approximately 1-dimensional curves in a 2-to-6
  dimensional space (depending on the number of available phase-space
  dimensions). Rather than constructing a data-ordering in real (e.g. position
  and velocity) space, we can instead find a 1-D subspace in which the stream is
  approximately a line, and the ordering is more evident. Following
  \autoref{sub:great_circle_coordinate_frame}, in
  \autoref{sub:self_organizing_maps} we describe a robust and performant means
  to recover an approximate 1-D subspace containing the stream and then order
  the stream data.

  \subsection{Great-Circle Coordinate Frame}
  \label{sub:great_circle_coordinate_frame}

    Constructing the stream-oriented coordinate system requires three
    parameters, a new on-sky origin -- at longitude $\ell_c$ and latitude $b_c$
    -- and a rotation angle about the origin. The new origin point is chosen to
    be the location of the progenitor, if known, and anywhere reasonable on the
    stream otherwise. In simulations the location of the progenitor is always
    known, even if the progenitor has fully dissolved. The rotation angle
    $\theta$ must be fit. Since errors in $\ell$, $b$ are negligible in both
    (contemporary) observations and simulations, the best-fit parameters are
    found by simple ensemble minimization of $\phi_2$ of the stream stars. For
    the fixed origin point ($\ell_c$, $b_c$) on the sky and rotation $\theta$,
    the transformation\footnotemark to a stream-oriented great-circle frame
    ($\phi_1, \phi_2$) is given in a Cartesian basis by \citep{Bovy2011}:
    \begingroup\makeatletter\def\f@size{10}\check@mathfonts
    \def\maketag@@@#1{\hbox{\m@th\large\normalfont#1}}%
    \setlength\arraycolsep{2pt}
    \begin{align}
      R &= \phantom{\cdot} \left[\begin{matrix}
        1 &   &   \\
          & - \cos{\left(\theta \right)} & \phantom{+} \sin{\left(\theta \right)} \\
          & - \sin{\left(\theta \right)} & \phantom{+} \cos{\left(\theta \right)}
      \end{matrix}\right]
      \cdot
      \left[\begin{matrix}
          - \cos{\left(b_c \right)} &   & - \sin{\left(b_c \right)} \\
            & 1 &   \\
          \phantom{+} \sin{\left(b_c \right)} &   & - \cos{\left(b_c \right)}
      \end{matrix}\right]
      \\
      &\phantom{=} \cdot
      \left[\begin{matrix}
        \cos{\phantom{+} \left(\ell_c \right)} & - \sin{\left(\ell_c \right)} &   \\
        \sin{\phantom{+} \left(\ell_c \right)} & \phantom{+} \cos{\left(\ell_c \right)} &   \\
          &   & 1
      \end{matrix}\right] \nonumber
    \end{align}
    \endgroup
    \footnotetext{ In \trackstream{}, the rotation transformation is implemented
        to be consistent with \astropypkg's \astropySkyOffsetFrame{} and the
        resulting frame can be used with the \astropypkg{} package
        \citep{Astropy2022}. }

\begin{figure}
      \script{method_frame-fit.py}
      \centering
      \includegraphics[width=1\linewidth]{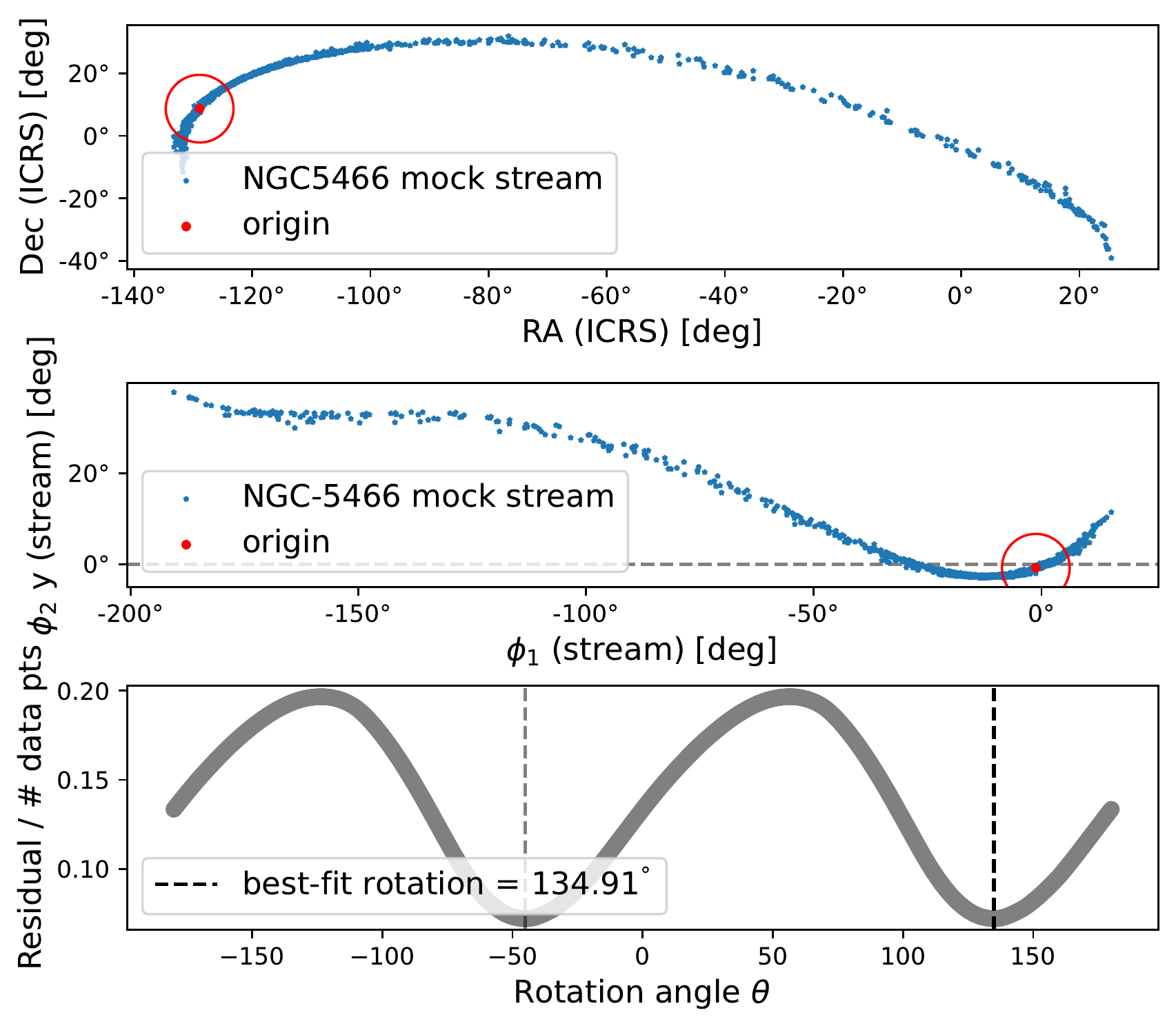}
      \caption{ Simulation of \stream{NGC 5466} \citep{El-Falou2022}.
        [top] \stream{NGC 5466} in an ICRS reference frame.
        [middle] Residual of
        $\frac{1}{N}\sum_{i=1}^{N}\left({\norm{\phi_2^{(i)}} - 0}\right)$, the
        longitude coordinate of \stream{NGC-5466} in a rotated reference frame,
        with rotation angle $\theta$ and origin set in top panel.
        [bottom] \stream{NGC-5466} in the best-fit rotated reference frame, with
        $\theta = 167^\circ$. The rotation angle has a $180^\circ$ degeneracy
        and the selected angle depends on the starting point of the minimizer. }
      \label{fig:method_frame-fit}
    \end{figure}

    \autoref{fig:method_frame-fit} shows an example construction of a rotated
    reference frame for a mock stream of \stream{NGC 5466} \citep{El-Falou2022}.
    The mock stream was simulated with the particle-spray code
    \galpystreamspraydf{} \citep{Qian2022}, now found in \galpy{}
    \citep{Bovy2015} but originally from \package{streamtools}
    \citep{Banik2009}, which is an implementation of the \citet{Fardal2015}
    method. The current position and velocity of the progenitor globular cluster
    is from \citet{Vasiliev2019}. The initial conditions (the progenitor before
    it is disrupted and forms tidal tails) are set by back-integrating the
    current phase-space position 5 Gyr in \galpyMWPotential{} \citep{Bovy2015}:
    a good approximation of the Galactic potential \citep{Bovy2016cx} composed
    of a linear combination of a spherical potential from a power-law density
    with an exponential cut-off Galactic bulge, an NFW dark matter halo
    \citep{NFW1996}, and a \citet{Miyamoto1975} disc. The initial progenitor is
    forward integrated in the same potential, forming the tidal tails shown in
    the top panel of \autoref{fig:method_frame-fit}.
    The stream-specific coordinate frame is the first step to ordering the data
    along the stream's path. Depending on the stream and projection effects,
    ordering by $\phi_1$ might be close (or quite far) to this correct ordering.
    Generally, ordering by $\phi_1$ is excellent for $\norm{\phi_1} \approx 0$,
    and progressively worse at larger $\norm{\phi_1}$.
    \autoref{fig:method_frame-fit} demonstrates the usefulness of constructing a
    stream-specific reference frame: in ICRS coordinates the Declination
    $\delta$ is not a one-to-one function of the Right Ascension $\alpha$, but
    the middle sub-figure shows the stream is very well ordered by $\phi_1$.
    Better yet, since \stream{NGC-5466} is reasonably thin and straight (in
    projection), the $\phi_2(\phi_1)$ ordering remains valid for the full
    observed length of the \stream{NGC-5466} mock stream.

    For long streams that wrap around the sky (called phase wrapping) or that
    curve such that $\phi_2(\phi_1)$ is not one-to-one, this ordering fails
    without further intervention. While this failure does not quite happen with
    \autoref{fig:method_frame-fit}, extrapolating from \stream{NGC-5466}, at
    $\phi_1 \gtrsim 50^\circ$ projection effects will ruin the simple
    $\phi_1$-ordering approach. In general, streams cannot be ordered well by
    their longitude coordinate. We require a coordinate that better tracks the
    structure of the stream. Looking at \autoref{fig:method_frame-fit}, finding
    the on-track coordinate is somewhat analogous to deforming the $\phi_1$
    dimension, wrapping it to meet the curvature of the stream. Another way of
    understanding the transformation is to consider again a galactocentric great
    circle. Intuitively we see that such an arc only projects to a line for an
    observer on the plane of the great circle. For all other locations, only a
    non-linear projection is guaranteed to map the curve to a line. In essence,
    the non-linearity of the projection encodes the spatial components of the
    galactocentric-to-observer reference frame transformation. We note, however,
    that this approach cannot easily be reverse engineered to provide a novel
    measure of the galactocentric frame parameters. Next we describe how we can
    find a non-linear projection to recover the 1-D subspace of a stream and use
    the on-stream coordinate to order the data.


  \subsection{Self-Organizing Maps} \label{sub:self_organizing_maps}

    To determine a stream's 1-D subspace we use the simple neural network method
    of Self-Organizing Maps (SOMs). SOMs are primarily for low-dimensional,
    discrete representations of data. Using prototype vectors linked in a
    lattice structure of a desired topology, SOMs approximate the data by
    iteratively updating the prototypes to approximate the data distribution. In
    final form, each prototype maps nearby high-dimensional data to a
    lower-dimensional lattice. For a detailed discussion we refer the reader to
    \cite{DataScience}.

    \subsubsection{The Algorithm, in 1-D.} \label{ssub:som_math_overview}

      Here we review the standard SOM algorithm and mathematics as they apply to
      a reduction from the higher-dimensional phase space of the positions (and
      velocities) to a one-dimensional subspace. The SOM can be extended to
      arbitrary features, perhaps to include photometric information, but we
      defer that for future work.

      The data consists of $N$ tracers measured at positions $\mbf{w}^{(n)}$ in
      the $D$-dimensional phase-space. In simulations, streams are generally
      best represented in galactocentric Cartesian coordinates, but observations
      are taken in spherical coordinates: two angular positions, possibly a
      radial distance, and corresponding kinematics. An important part of the
      SOM is the data-space distance metric, which is a complication when
      working in Cartesian coordinates versus on the surface of a sphere. For
      spherical coordinates the correct distance metric is the
      \citet{Vincenty1975} formula; this is used in \trackstream{} and the
      figures in this paper. For many streams the angular separation between
      tracers is small, so it is reasonable to work within a flat sky
      approximation and simplify to a Euclidean distance metric in Cartesian
      coordinates. Therefore, for simplicity here we only discuss the method in
      Cartesian coordinates.

      The SOM will learn the structure of the data using a smaller number $K$ of
      ``prototype'' vectors $\mbf{p}^{(k) \in [1, K]}$ in the same space as the
      data. The need for only a few prototype vectors is analogous to how only
      two points are required to describe a line, no matter how many data points
      are sampled from the line. The relationship between the $\mbf{p}_{k}$
      prototype vectors determines the topology and ordering of the data
      $\mbf{w}^{(n)}$. Returning to the line example, the relationship between
      the two line points is a linear segment, so when fit to a dataset drawn
      from a line distribution, the SOM should describe the data well. 

      Generically the prototypes' relationships are described with a feature map
      $\mbf{Q}$, where each prototype $\mbf{p}^{(k)}$ has a corresponding
      lattice node $\mbf{q}_k$. However, since streams are one-dimensional the
      only pertinent information is the index: $\mbf{q}_k \equiv k$. $\mbf{Q}$
      encodes that prototypes $\mbf{p}^{(i)}$, $\mbf{p}^{(j)}$ are neighbors if
      and only if the corresponding lattice points $\mbf{q}_i$, $\mbf{q}_j$ are
      neighbors -- i.e. $|i-j| = 1$.

      With this review of feature and lattice space we are ready for the SOM
      algorithm:

      \label{som_algorithm_steps}
      \begin{enumerate}[label={(\bfseries \arabic*):}, leftmargin=*]
      \itemsep0.5em
        \item \tbf{Given} the data in the stream coordinates
              (\autoref{sub:great_circle_coordinate_frame}),
        \item \tbf{Choose} $K$ prototypes $\mbf{p}^{(k)}$ with linear lattice
              $\mbf{Q}$.
        \item \tbf{Iteratively},
          \begin{enumerate}[label={(\bfseries \arabic*):}, leftmargin=*]
          \itemsep0.5em
            \item Select the next datum $\mbf{w}^{(n)}$.
            \item Update the data-space position of the nearest prototype,
            keeping the lattice unchanged.
            \item Similarly update all the prototypes' topological (lattice)
            neighbors, making smaller updates for more distant prototypes.
          \end{enumerate}
      \end{enumerate}

      By multiple iterations the prototypes will be updated to lie near the data
      in position space and capture the structure of the data in lattice space.
      For a more mathematically detailed explanation see Appendix
      \ref{app:som_math} and \trackstream{} for a code implementation.


    \subsubsection{Projecting on the SOM} \label{sub:som_projection}

      After applying the SOM and deriving the set of prototype vectors, the next
      step is to project the data into the 1-D space and order the data in a
      manner consistent with the topology. There are numerous projection
      algorithms, such as U-matrix maps \citep{Ultsch1990}, that are intended
      for clustering data by the lattice-point distance. The $\mbf{Q}$ is not of
      interest; we want to project on the real-space structure of the SOM, and
      so implement a custom algorithm to connect and project onto the SOM
      prototypes. While it is tempting to connect the SOM prototypes with a
      smooth curve, e.g. a Bezier curve \citep{Bezier1982} or 3rd order
      piece-wise polynomial spline, not only is projection onto a curve
      challenging, it is not correct for an SOM with a linear lattice. The SOM
      topology is a linear structure and the SOM prototypes are connected by
      line segments. Projecting each datum reduces to the task of 1) identifying
      the correct line segment and 2) projecting onto that segment.

\begin{figure*}
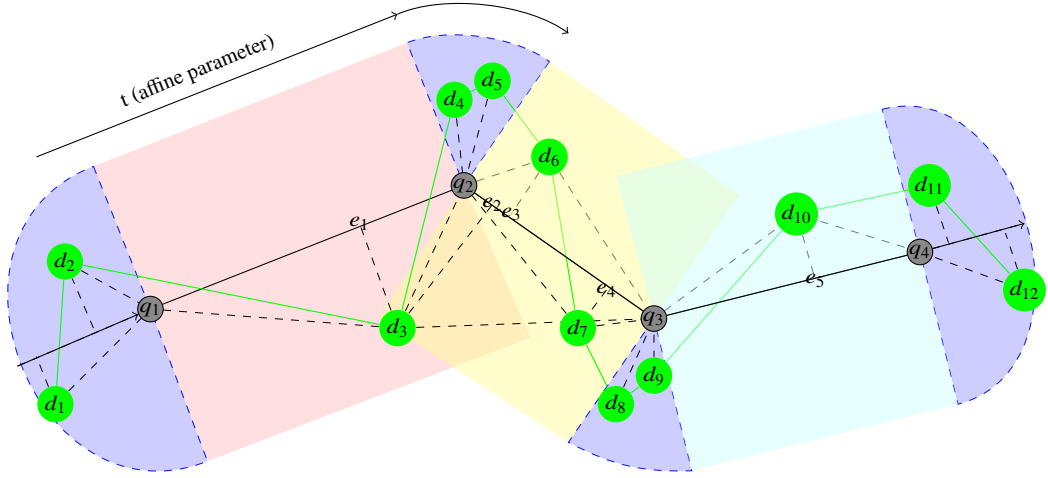

        \centering
        \ctikzfig{static/som_tikz}
        \caption{
          \emph{Illustration of ordering data from SOM lattice nodes.}
          The SOM creates a 1-D lattice of connected nodes ($q$'s, gray) ordered
          by proximity to the designated origin, then along the lattice. Data
          points ($p$'s, green) are assigned an order from the SOM-lattice. The
          distances from the data to each node are computed. Likewise the
          real-space projections are found for each data point on the edges
          connecting each SOM node. All projections lying outside the edges
          (shaded regions) are eliminated. Also eliminated are all but the
          closest nodes. Remaining edges and node connections are in dotted
          block, with projections labeled $e$. Data points are sorted into the
          closest node regions (blue) and edge regions (shaded). Data points in
          end-cap node regions are sorted by extending the projection onto the
          nearest edge. Data points in edge regions are ordered by projection
          along the edge. Data points in intermediate node regions are ordered
          by the angle between the edge regions. }
        \label{fig:som_index_algorithm}
      \end{figure*}

      The custom algorithm, with examples drawn from
      \autoref{fig:som_index_algorithm}, is as follows:
      \begin{enumerate}[label={(\bfseries \arabic*):}, leftmargin=*]
      \itemsep0.5em
        \item Get the ordered lattice points (black) from the SOM. The first
        point is the nearest to the specified origin (e.g. the stream
        progenitor, see \autoref{sub:great_circle_coordinate_frame} for details)
        and subsequent points draw their ordering from the topology.
        \item Compute the directed real-space segments from one lattice point to
        the next.
        \item For each data point, compute the real-space projection onto each
        segment
            \begin{align}
              \mbf{r} &= \mbf{p}^{(k)} + \ t \cdot (\mbf{p}^{(k+1)} - \mbf{p}^{(k)}),
              \\
              t &= \frac{(\mbf{w}^{(n)}-\mbf{p}^{(k)}) \cdot (\mbf{p}^{(k+1)}-\mbf{p}^{(k)})}{|\mbf{p}^{(k+1)}-\mbf{p}^{(k)}|^2}
            \end{align}
            for data point $\mbf{w}^{(n)}$, and lattice points $\mbf{q}_k$.
            Projections between the two lattice points have $0<t<1$.
        \item For each data point, compare the distance to each node and to each
        projection to find which node / segment it should be associated with.
          \begin{enumerate}[label={(\bfseries \arabic*):}, leftmargin=*]
          \itemsep0.5em
              \item data with $t=0,1$ are closest to a node.
              \item data with projection parameter $0<t<1$ will be closer to
              that projection than any node (e.g. $d_3$) and can be
              distinguished between segments by the point-to-projected-point
              distance. E.g., data point (green) $d_3$ is closer to projected
              point $e_1$ than to $e_2$ and therefore belongs in the $q_1$-$q_2$
              segment.
              \item data with $t < 0$ or $1 < t$ and that are closest to a
              lattice terminal node are within the ``end-caps'', e.g, $d_1$.
              \item data with $t < 0$ or $1 < t$ and not in an end-cap are in a
              convexity, e.g, $d_5$.
          \end{enumerate}
        \item Within each ``block'' (end-cap, segment, convexity) organize the
        associated data points:
          \begin{enumerate}
              \item data in an end-cap are organized by distance along the
              projection (extending $t < 0 \ \& \ 1 < t$).
              \item data in a segment are ordered by $t$.
              \item data in a convexity are ordered by their angular coordinate
              sweeping from one segment to the next. This is why $d_8$ precedes
              $d_9$ in \autoref{fig:som_index_algorithm}.
          \end{enumerate}
        \item Order the blocks by the lattice node ordering.
      \end{enumerate}

\begin{figure}
        \script{method_som-fit_solar-circle.py}
        \centering
        \includegraphics[width=1\linewidth]{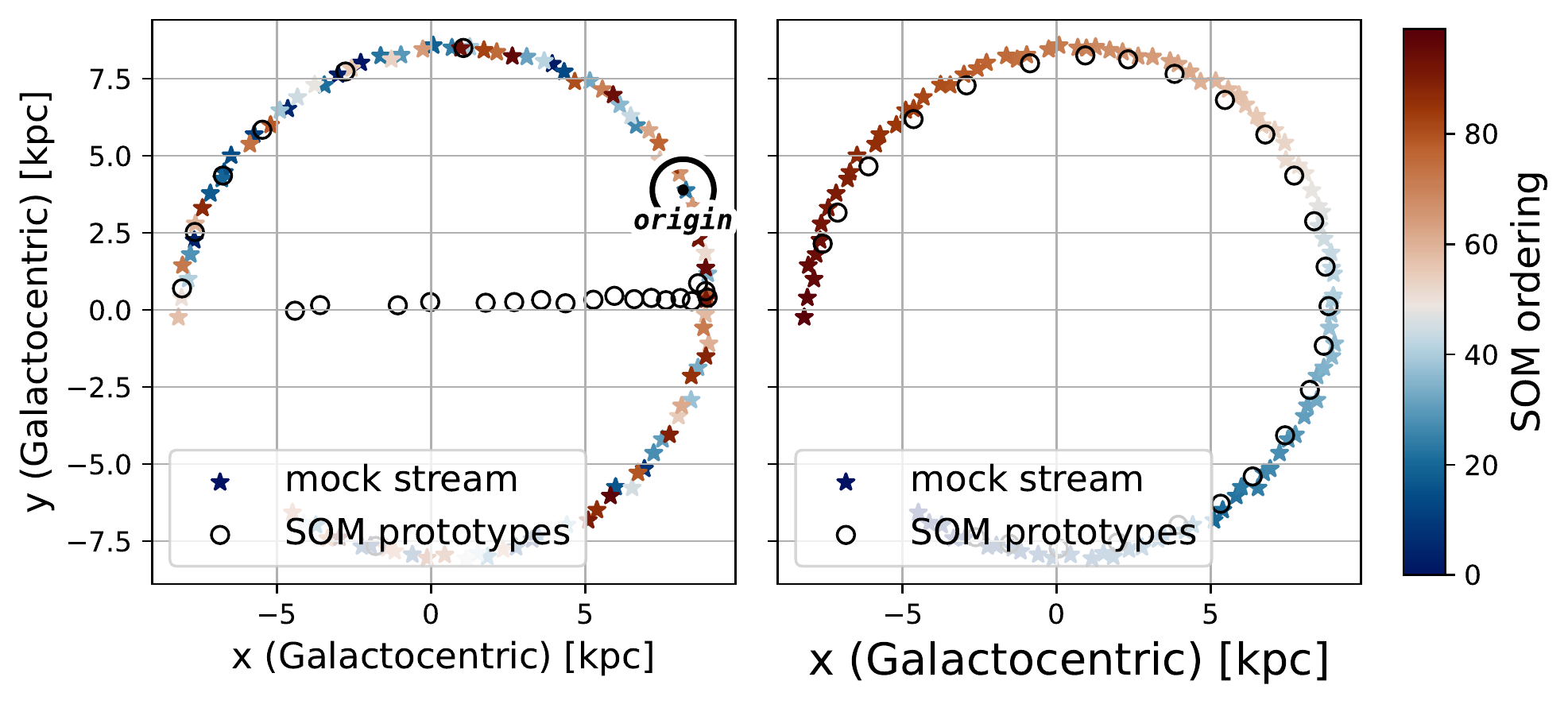}
        \caption{
          \emph{The SOM-derived ordering of a mock stream.}
          The left shows a z-projection of the stream, with stars "observed" in
          a random sequence. The initial position of the SOM prototypes (gray)
          do not match the data. The right figure is the same but with the
          trained SOM-derived ordering. The trained SOM prototypes are included
          (black circles), showing how SOMs require only a relatively small
          number of prototypes to capture the structure of the data. }
        \label{fig:method_som-fit_solar-circle}
      \end{figure}

      In short, the SOM is trained on the data and then used to order the data
      by projecting the data onto the SOM's 1-D subspace. Standard methods to
      order a stream rely on the stream being a one-to-one function in an
      observational coordinate, for example the position angle $\phi_1$ on the
      sky. \autoref{fig:method_som-fit_solar-circle} shows that SOMs are not
      subject to this limitation and can be used to order streams that wrap
      around the Milky Way or are in other ways not one-to-one functions in any
      observational coordinate. \autoref{fig:method_som-fit_solar-circle} is a
      mock dataset of stars in a stream-like structure, created by integrating
      the orbital path of the sun \citep{Reid2004, GRAVITY2018, Drimmel2018,
      Bennett2019} in \galpyMWPotential{} for 200 Myrs. Stars are drawn randomly
      along the ``stream`` from a $\sigma=100 \; [\rm{pc}]$ Gaussian
      distribution. The SOM prototypes are initialized using equi-frequency
      binning. Unlike the \stream{NGC 5466} simulation (see
      \autoref{fig:method_frame-fit}), the binning produces a poor match to the
      data. However, after $10^5$ iterations (at $\sim\!18,000$
      iterations/second on a 2018 2.7GHz i7 MacBook Pro) the SOM fits well and
      orders the data well. Note that most of the trained SOM prototypes do not
      lie directly on the mock stream. For a tight curve like this one, the
      offset is expected: points near to a prototype $p_k$ influence its
      location, so prototypes are generally interior to a convexity as that is
      locally the best average position.

      \begin{figure}
        \script{method_som-fit.py}
        \centering
        \includegraphics[width=1\linewidth]{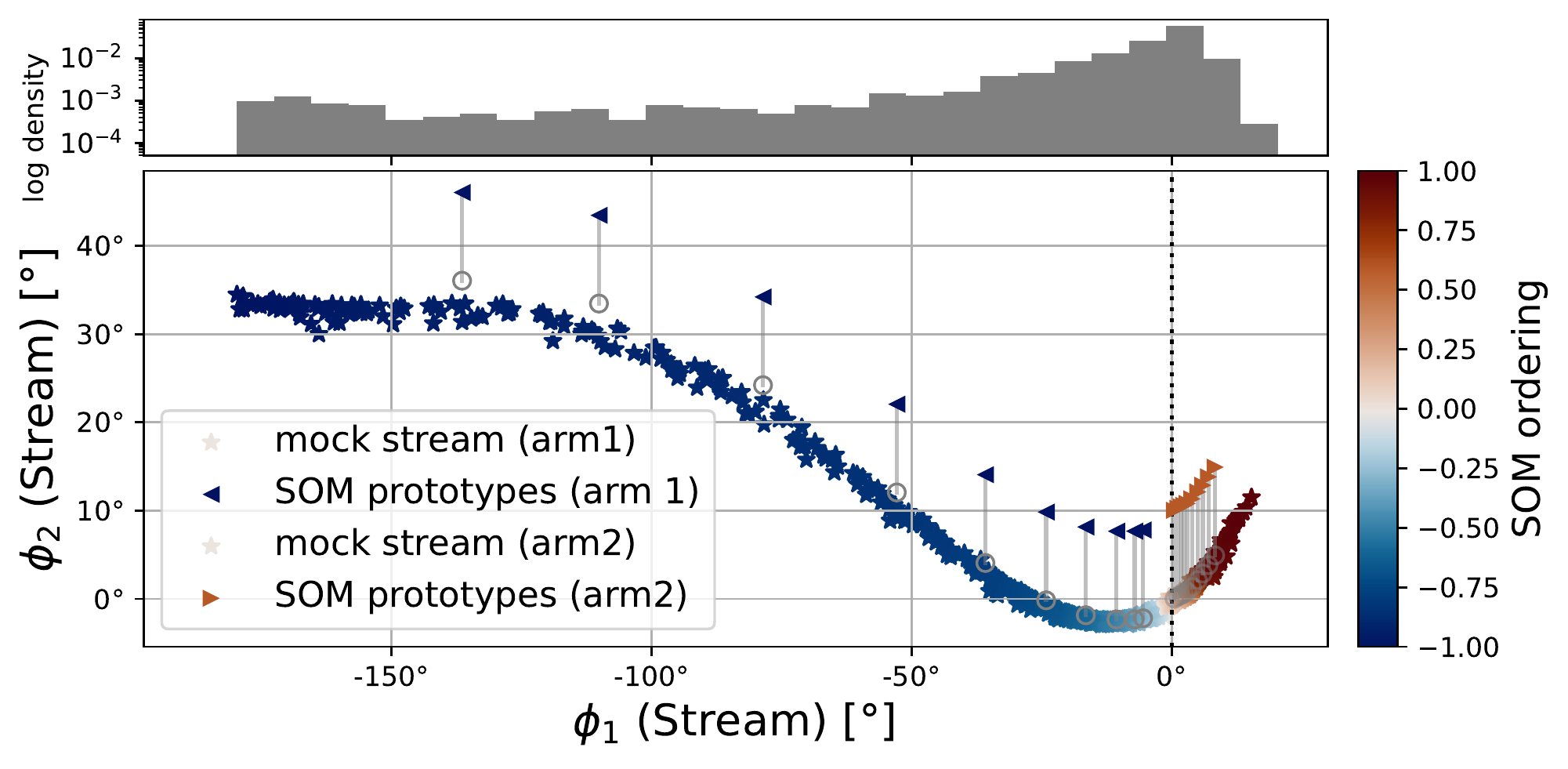}
        \caption{ \emph{Simulation of \stream{NGC-5466} ordered by SOM.} The
          \stream{NGC-5466} simulation is the same as
          \autoref{fig:method_frame-fit}. The best-fit frame is determined by
          the method described in \autoref{sub:great_circle_coordinate_frame}
          (and shown in \autoref{fig:method_frame-fit}). The SOM has 10
          prototypes (per arm), initialized by equi-frequency binnning of the
          $\phi_1$ data (shown in the top histogram) and trained for $10^4$
          iterations before being used to order the data. The trained prototypes
          are shown offset from their actual location. In regions of very high
          density the prototypes lie on the stream, while in low density regions
          they do not. This is a consequence of local averaging and shows the
          SOM is correctly responding to the structure of the data. The SOM is
          used to order the data using the process described in
          \autoref{sub:som_projection}. }
        \label{fig:method_som-fit}
      \end{figure}

      \autoref{fig:method_som-fit} shows the SOM-derived order for a simpler
      example, the mock \stream{NGC-5466} stream also shown in previous figures.
      Two SOMs are used, one for each stream arm. Each SOM is trained for 10,000
      iterations ($\sim0.5$ second on a 2018 2.7GHz i7 MacBook Pro). The
      prototypes cluster more densely where the data are dense, such as in the
      right-hand arm, but still provide good coverage of the whole stream. Shown
      by the colormap, the SOM order is good, with stars near in position
      likewise near in the ordering. For this stream the SOM is not strictly
      required as the position angle $\phi_1$ is sufficient to order the data.
      The SOM can be initialized with random prototypes and a connecting
      lattice, or with an informed guess. Given enough iterations both will
      typically converge to the same solution, but the former will require many
      more iterations than the latter. Rather than skipping the SOM step, we
      instead use an informed initial condition\footnotemark and reduce the
      training iterations, e.g. only the 10,000 used here. For
      \autoref{fig:method_som-fit} the SOM prototypes are initialized by
      equi-frequency binning in $\phi_1$.

      \footnotetext{In \trackstream{} any initialization may be used. If none is
      specified the code defaults to equi-frequency binning, placing the a
      prototype at the average location of every $n$th set of data points, where
      $n$ is the number of data divided by the number of prototypes.
      Equi-frequency binning equitably spreads the prototypes over the data, if
      the data is one-to-many in $\phi_1$. If the data is many-to-one in
      $\phi_1$, for example with phase-wrappings, the binning is poor.}
    
      Were \stream{NGC-5466} simulated again in a slightly perturbed potential
      or with similar initial conditions, the stream would be generally similar.
      A SOM trained on the previous simulation can be the initial condition for
      the SOM of the new simulation, therefore requiring very few iterations to
      converge and match the new simulation. Re-using an SOM for a similar
      simulation is a technique that works for any stream, but is particularly
      useful for complex streams, for example Sagittarius which wraps several
      times around the Galaxy \citep{Koposov2012}. Consider a generative model
      run as part of a likelihood calculation in a Markov Chain Monte Carlo
      (MCMC). Steps in the MCMC are to nearby points in the parameter space of a
      potential and progenitor, and as discussed in the context of
      \stream{NGC-5466}, the  generated mock stream will be similar to the
      previous stream. For the initial state of the MCMC the SOM must either run
      for many iterations or have an informed starting point. For subsequent
      steps, the previous SOM may be used as an informed starting point and
      should require only a little tuning, not wholesale retraining. Generally,
      we expect the generative model of the stream to be the time-limiting
      portion of any pipeline, not the SOM.





\section{Methods: Constructing Stream Track by Kalman Filter}
\label{sec:kalman_filter}

  Cold streams are approximately 1-D and using a SOM we have shown that stream
  data can be projected into a subspace that is better suited for ordering than
  on-sky coordinates. The remaining task is to move through the ordered data,
  fitting the track, width, and any other observables along the stream. We will
  be using a first-order Newtonian Kalman filter to fit the phase-space track.
  Before deriving how Kalman filters work, it is beneficial to take a step back
  and outline the challenges of the data and take stock of alternative
  approaches.


    The data are comprised of observations (real or simulated) of stars known to
    be members of the stream. This data has at least the on-sky position, and
    also perhaps the distance and kinematics of each star. Consider a dataset of
    only the on-sky positions. For finding the on-sky stream path it is
    appealing to fit with a low-order polynomial. For a sufficiently trivial
    stream, this approach can work. However, if the stream has any kink(s) --
    i.e. from interactions -- or curvature -- as can happen from projection into
    an observer's reference frame -- then the stream path might not be a
    function, as can be seen in \autoref{fig:method_som-fit_solar-circle}, as is
    required to fit a low-order polynomial.    

    To achieve a general solution the data cannot be fit one coordinate against
    another; much better is to have an affine variable and instead fit
    parametrically for each coordinate. For a stream, this affine parameter can
    be the arc-length, i.e. the angular distance along the mean path of the
    stream. The point-to-point distances of the data projected onto the SOM's
    subspace are very nearly this arc-length. With the arc-length as the affine
    parameter one might do a Gaussian process regression \citep{Cervone2015} or
    fit any manner of function: low-order polynomial, B-spline, Bezier curve.
    However, all these procedures are susceptible to the same problem, the SOM
    is an unsupervised machine learning technique and is not guaranteed to
    converge. In particular, if trained quickly it can make small mistakes in
    the ordering. In principle this shortcoming can be overcome by more training
    iterations, but if time is a constraint, this extra step may not be
    practical and may require human supervision to confirm an optimal fit, which
    is likewise impractical. Therefore, a general fitting procedure for stream
    data should on the one hand use the SOM-derived data ordering as an
    approximate affine variable, but also be insensitive to small inaccuracies
    in that ordering. A fitting method that meets both these requirements is the
    Kalman filter, presented here.



    The Kalman filter is a well-established algorithm to estimate a joint
    probability distribution over unknown variables from a time series of
    measurements that may have statistical noise \citep{Kalman1960}. In this
    context the unknown variables characterize the stream path and the
    measurements are the 2-to-6 astrometric dimensions of the stream's stars.

    Kalman filters fit data in a very different manner than e.g. polynomial
    fitting, which works by minimizing some global residual function. Instead,
    Kalman filters start at some initial state (a ``position'' in the
    astrometric phase-space) and sequentially introduce measurements, evolving
    the state and associated uncertainties according to a dynamical model of the
    system. This method is a two-step process. The first step is \textit{a
    priori}: predicting from the prior state and the dynamical model the current
    state variables and uncertainties. The next step is \textit{a posteriori}:
    adjusting the predicted state given the incoming measurement. The posterior
    state estimate is a weighted average of the prior and measurement, with the
    weight determined from both the prior and measurement uncertainties. A
    low-noise measurement against an uncertain prior will favor the measurement
    in the posterior state estimation, and vice versa for a noisy measurement
    and high-confidence prior.



    One of the simplest dynamical models, and the one used in this work, is a
    first-order Newtonian model. In this filter states are evolved according to
    $\mbf{x} \mapsto \mbf{x} + \mfrak{v} \Delta t$, where $\mbf{x}$ is the
    position of an imaginary point moving along a smooth differentiable manifold
    passing near the positions of the stars. Here, $\mfrak{v}$ is a
    pseudo-velocity of the imaginary point. A more complex dynamical model might
    incorporate the Galactic potential, taking into account the acceleration
    field. However, incorporating the potential introduces a lot of complexity
    to avoid poor physics, for negligible gain. Adjacent stars on a stellar
    stream do not move along the same orbit, thus simply evolving the state in
    the potential is incorrect. For example, the streak-line model
    \citep{Kupper2012} back-integrates the progenitor then forward-integrates a
    stream star's orbit from its progenitor ejection time. A good dynamical
    model might use a similar approach for the next-state prior: back-integrate
    the current state to its ejection (call this $\Delta t_{\rm{eject}}$),
    forward integrate the progenitor for the set time step, then forward
    integrate an ejected star for the same $\Delta t_{\rm{eject}}$. The
    first-order Newtonian model is computationally faster for equally good
    results and does not require information about the potential.

  \subsection{Point-to-point times} \label{sub:distances_as_times}

    Time-series methods like the Kalman filter require a time step, however the
    stream data are not a time series (see \autoref{sec:ordering_the_data}):
    first, the observations are not of the same object, and second, the
    observations are not taken in any relevant time order. Despite that the data
    are not a true time series, we can treat the data as if it were time ordered
    and amenable to the Kalman filter. By using the frame-rotated
    (\autoref{sub:great_circle_coordinate_frame}), SOM-ordered stream
    (\ref{sub:self_organizing_maps}) the data are organized by distance along
    the as-yet-unknown stream track. Consider moving along the track, passing by
    each data-point with some time variable $t$: the distance ordering is
    ordinally equivalent to the time ordering, index-by-index.

\begin{figure}
      \script{method_time-proxy.py}
      \centering
      \includegraphics[width=0.9\linewidth]{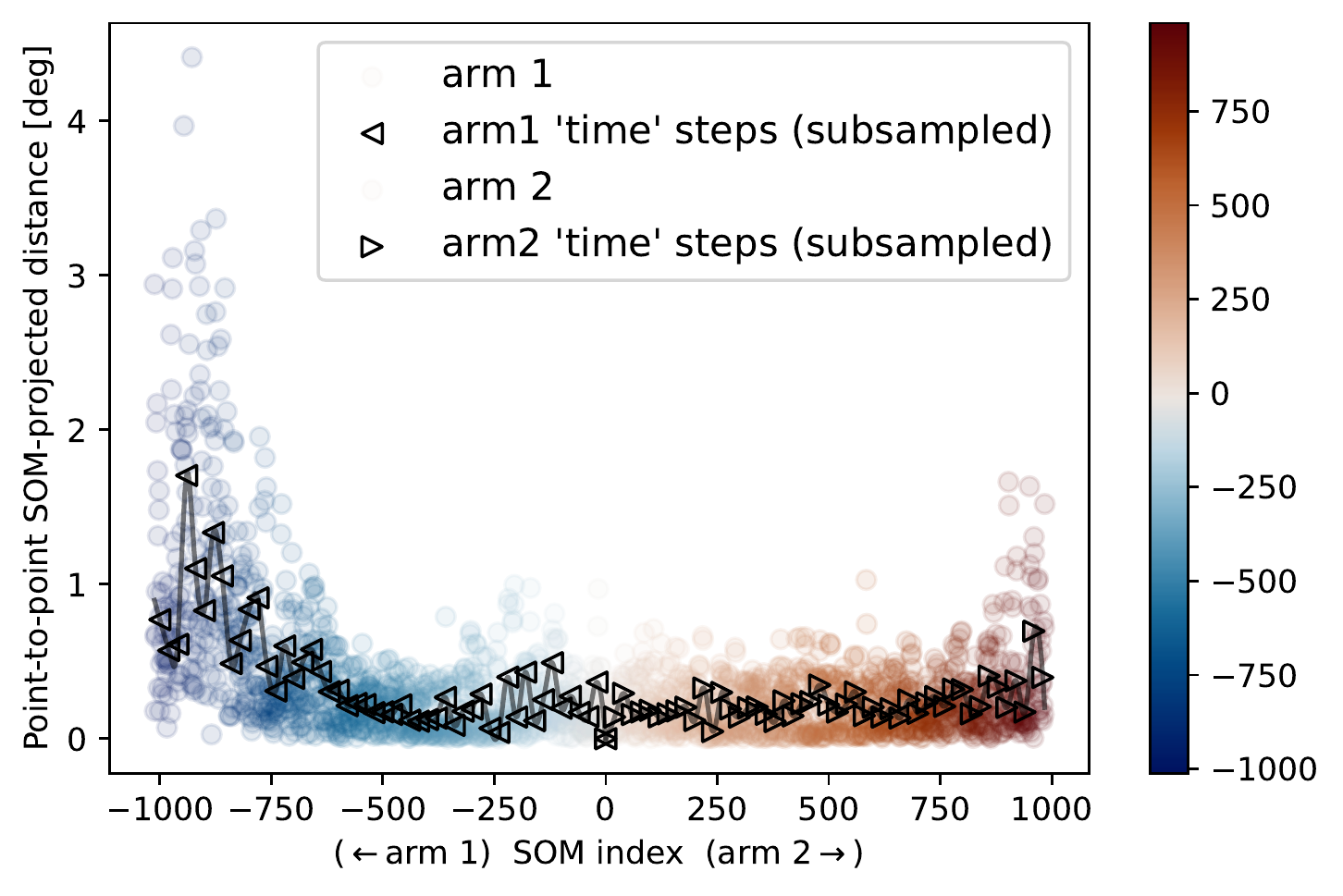}
      \caption{\emph{Point-to-point SOM-projected distance as proxy for the
      Kalman filter time step.} After the data of \stream{NGC-5466} are
      organized by the SOM (see \autoref{fig:method_som-fit}), the data are
      projected into the 1-D subspace of the SOM. The point-to-point distance is
      a good measure of the distance the Kalman filter should travel when
      determining the location of the next point on the stream track.
      Interpreting the distance step as a time step both enables the use of the
      Kalman filter and also helps it closely follow the stream. As there is
      large variation in the projected distance, the time steps are smoothed.}
      \label{fig:method_time-proxy}
    \end{figure}

    To re-interpret distances as times we take the time step $\Delta t$ as the
    point-to-point distance in the SOM projection of the data. For the
    \stream{NGC 5466} stream this distance is shown in
    \autoref{fig:method_time-proxy} as the colored points, indexed by the SOM
    ordering. We then smooth the time-step by kernel convolution:
    \begin{equation}
      \Delta{t}_{i} = \rm{\mathit{w}[n](i|\Delta X)},
    \end{equation}
    where $\mathit{w}[W](i|X)$ is the window average of the point-to-point
    distances in data $X$, of window size $W$, centered on the $i$th index. The
    choice of window is arbitrary, with long-tailed windows more closely
    matching $\Delta{t}$ than short-tailed windows, which will better reflect
    the local density. We use a Dirichlet window (i.e. a moving window average)
    of window-width six -- three points per side. The smoothing reduces the
    effects of large outliers and the result is not sensitive to larger window
    sizes, so this small size is kept fixed. The smoothed steps are shown as
    connected, directed black triangles in \autoref{fig:method_time-proxy}. The
    variation in the smoothed steps is much smaller than the unsmoothed data.
    Now the ``time''-step is tuned so that $v \Delta{t} \approx \Delta{d}$. The
    velocity $\mfrak{v}$ of the Kalman Filter is not affected as we change only
    the time-step to make the filter accurately reach each data point.

  \subsection{Running the Kalman Filter} \label{sub:filter_setup}

    In this section we set up a Kalman filter and discuss how it is run. For a
    detailed review, \citet{KalmanFiltersinPython} offers an excellent
    discussion of Kalman filters and how to implement them in \python.
    \citet{KalmanFiltersinPython} works in Cartesian coordinates, but the
    methods are adaptable to spherical coordinates by using appropriate distance
    measures and being careful of angular coordinate wrappings. \trackstream{}
    supports both Cartesian and spherical metrics, which are used where
    appropriate for the figures. Here, for simplicity, we only present the
    Kalman filter math in Cartesian coordinates.

    To set up a model, we assume that the stars are in a vicinity of a
    one-dimensional differentiable manifold following the stream track,
    parameterized as $\mbf{c}(t)$, where $t$ represents an artificial time, such
    as the scaled arc-length. Thus, the actual positions of the stars, indexed
    according to the SOM ordering, are given by the model
    \begin{equation}
        \mbf{w}^{(k)} = \mbf{c}(t_k) + \mboxbf{\xi}^{(k)},
    \end{equation}
    where $\mboxbf{\xi}^{(k)}$ is an unknown offset parameter. To link the model
    to observations, let $\mbf{w}^{(k)}_{\rm obs}$ denote the observed position
    of the $k$th star, representing the data,
    \begin{equation}\label{eq:observation}
        \mbf{w}^{(k)}_{\rm obs}
        = \mbf{c}(t_k) + \mboxbf{\xi}^{(k)} + \mboxbf{\epsilon}^{(k)}
        = \mbf{c}(t_k) + \mbf{n}^{(k)},
    \end{equation}
    where $\mboxbf{\epsilon}^{(k)}$ is the observation error. 

    To set up the probabilistic phase-space model, we define an extended
    variable,
    \begin{equation}
        \mbf{x}^{(k)} = \begin{bmatrix} \mbf{c}^{(k)} & \mfrak{v}^{(k)} \end{bmatrix}^T,
    \end{equation}
    where $\mbf{c}^{(k)} = \mbf{c}(t_k)$, and $\mfrak{v}^{(k)}$ is the
    artificial velocity vector with respect to the parameter $t$, defined as
    $\mfrak{v}^{(k)} = \dot{\mbf{c}(t_k)}$. A local first-order Newtonian model
    gives us a propagation model,
    \begin{align}\label{eq:propagation}
        \mbf{c}^{(k)} &= \mbf{c}^{(k-1)} + \Delta t_{k} \mfrak{v}^{(k-1)} + \mbf{m}^{(k)}_1,
        \\
        \mfrak{v}^{(k)} &= \mfrak{v}^{(k-1)} + \mbf{m}^{(k)}_2, \nonumber
    \end{align}
    where $\Delta t_{k} = t_{k} -t_{k-1} $, and  $\mbf{m}^{(k)}_1$ and
    $\mbf{m}_2^{(k)}$ represent the approximation error in the first order model
    for the position and the zeroth order model for the velocity propagation,
    respectively.
 
    The Equation \eqref{eq:propagation}-\eqref{eq:observation} pair is written
    as a propagation-observation pair for the phase space variable
    $\mbf{x}^{(k)}$ as
    \begin{align}
        \mbf{x}^{(k)} &= \mbf{F}_{k} \mbf{x}^{(k-1)} + \mbf{m}^{(k)}, \label{eq:kalman_forward1}
        \\
        \mbf{w}^{(k)}_{\rm obs} &= \mbf{H}_k \mbf{x}^{(k)} + \mbf{n}^{(k)}, \label{eq:kalman_forward2}
    \end{align} 
    where the propagation-observation matrices are given by
    \begin{equation}
        \mbf{F}_k =
            \begin{bmatrix}
                \bf{1} & \Delta t_k \bf{1}  \\
                {\bf 0} & \bf{1}
            \end{bmatrix}
            \quad \mbf{H}_k = \left[\bf{1},{\bf 0}\right],
    \end{equation}
    and  $\bf{1}$ and ${\bf 0}$ are a unit and zero $W\times W$ matrices, where
    $W$ is the number of phase-space dimensions.

    The propagation-observation pair allows us to use the Kalman filtering and
    Kalman smoothing to estimate the parameterized manifold points
    $\mbf{c}^{(k)}$. To set up a Gaussian model, we postulate that the state
    noise vectors $\mbf{m}^{(k)}$ and the observation noise $\mbf{n}^{(k)}$ are
    mutually independent and zero mean Gaussian random variables. The
    corresponding covariance matrices are set as
    \begin{equation}
        \rm{cov}\big(\mbf{m}^{(k)}\big)
        = \mbf{Q}_k
        = \lambda^2
            \begin{bmatrix}
                \frac{1}{4}\Delta t_k^4 \bf{1} &  \frac{1}{2}\Delta t_k^3 \bf{1} \\
                \frac{1}{2}\Delta t_k^3 \bf{1} & \Delta t_k^2\bf{1}
            \end{bmatrix},
    \end{equation}
    a commonly used form in tracking applications acknowledging the dependence
    of the first-order forward Euler process on the time step. The dependence
    that $\mbf{Q}_k$ represents is the error that a smooth function can be
    modeled with discrete steps, using only a velocity term to update those
    steps. Similar to how a first-order Taylor expansion has second-order
    errors, the first-order Newtonian model has second-order $\Delta{t}^2$ (and
    higher) errors from not accounting for the smoothness. In the above formula,
    $\lambda>0$ is a tuning parameter. For the observation error, we write
    \begin{equation}
        \rm{cov}\left(\mbf{n}^{(k)}\right)
        = \mbf{R}_k
        = \begin{bmatrix} \left(\sigma^{(k)}_{\rm stream}\right)^2 + \big( \sigma^{(k)}_{\rm meas}\big)^2 \end{bmatrix} \bf{1},
    \end{equation}
    where $\big(\sigma^{(k)}_{\rm stream}\big)^2$ accounts for the random
    variable $\mboxbf{\xi}^{(k)}$ represents the internal dispersion of the
    stars within the stream, and $\big( \sigma^{(k)}_{\rm meas}\big)^2$ is the
    variance of the observation error.
    
    The Kalman filtering algorithm is a sequential Bayesian filtering method,
    generating the conditional distribution of $\mbf{x}^{(k)}$ given all the
    past observations $\mbf{w}^{(j)}_{\rm obs}$, $1\leq j\leq k$. Under the
    assumption that both the model and observation noises are independent and
    Gaussian, the conditional distributions are also Gaussian, thus completely
    characterized by the conditional mean $\mbf{x}_{k\mid k}$ and covariance
    $\mbf{P}_{k\mid k}$, 
    \begin{equation}
        \pi\big(\mbf{x}^{(k)}\mid \mbf{w}^{(1)},\ldots,\mbf{w}^{(k)}\big)
        = \mcal{N}\big( \mbf{x}^{(k)}\mid \mbf{x}_{k\mid k}, \mbf{P}_{k\mid k}\big).
    \end{equation}
    $\mbf{x}_{k\mid k}$  and $\mbf{P}_{k\mid k}$ based on the previous state
    parameters $\mbf{x}_{k-1\mid k-1}$ and $\mbf{P}_{k-1\mid k-1}$ are updated
    in two phases, {\em prediction step} and {\em updating (or analysis) step}.
    The {\em a priori} prediction step is based solely on the propagation model,
    \begin{align}
        \mbf{x}_{k\mid k-1} &= \mbf{F}_k \mbf{x}_{k-1\mid k-1},
        \\
        \mbf{P}_{k\mid k-1} &= \mbf{F}_k\mbf{P}_{k-1\mid k-1} \mbf{F}_k +{\mathbf Q}_k.
    \end{align} 
    As the new observation arrives, the predicted distribution is considered as
    prior, and the update is obtained by Bayes' formula. Writing the pre-fit
    residuals of the mean and covariance as
    \begin{align}
        \mbf{y}_k &= \mbf{w}^{(k)}_{\rm obs} - \mbf{H}_k \mbf{x}_{k\mid k-1},
        \\
        \mbf{S}_k &= \mbf{H}_k \mbf{P}_{k\mid k-1} \mbf{H}_k^T + \mbf{R}_k,
    \end{align}
    the {\em a posteriori} update of the mean and covariance assume the forms
    \begin{align}
        \mbf{x}_{k\mid k} &= \mbf{x}_{k\mid k-1} + \mbf{K}_k \mbf{y}_k
        \\
        \mbf{P}_{k\mid k} &= \mbf{P}_{k\mid k-1} - \mbf{K}_k \mbf{H}_k\mbf{P}_{k\mid k-1},
    \end{align}
    where $\mbf{K}_k$ is the {\em Kalman gain} matrix, 
    \begin{equation}
        \mbf{K}_k = \mbf{P}_{k\mid k-1} \mbf{H}_k^T \mbf{S}_k^{-1}.
    \end{equation}
    The posterior state estimate is a weighted average of the prior and
    measurement, with the weight -- the {\em Kalman gain} -- determined from
    both the prior and measurement uncertainties. A low-noise measurement
    against an uncertain prior will favor the the measurement in the posterior
    state estimation.

    The algorithm requires the initial input $(\mbf{x}_0,\mbf{P}_0)$. To define
    the mean $\mbf{x}_{0\mid 0}$, we choose the mean position of a few  stream
    points closest to the origin, rather than the origin itself, since streams
    originate at the edge of their progenitor, not their center. The initial
    mean velocity is the normalized mean of the differences between the first
    few consecutive stream points. The covariance $\mbf{P}_{0\mid 0}$ is defined
    as a diagonal matrix, with the variances estimated by using the first few
    stream points. The approximation of $\mbf{P}$ as a diagonal matrix is
    corrected after the first time step by $\mbf{Q}_k$ and later, in a
    back-propagation step, discussed next.
    
    Kalman filtering, and other Bayesian filtering methods are based on the
    Markov models in which the next update depends on the past. When the process
    is run on-line, the data arriving sequentially as a time series, this
    updating approach is natural, while when all data are available as in the
    present case, no preferential direction of time can be justified. A common
    approach to find an estimate for the posterior distribution
    $\pi(\mbf{x}^{(k)}\mid \mbf{w}^{(1)},\ldots,\mbf{w}^{(D)})$, $1\leq k\leq
    D$, is to use Kalman smoothing, in which the filter is run in the reverse
    order, using the forward filtered distributions as priors.  Starting with
    the final distribution with mean $\mbf{x}_{D\mid D}$ and covariance
    $\mbf{P}_{D\mid D}$, the  the Rauch-Tung-Striebel (RTS) smoother
    \cite{RTS1965} algorithm walks backwards by computing the mean and
    covariance $\mbf{x}_{k\mid D}$ and $\mbf{P}_{k\mid D}$ through the updating
    formulas
    \begin{align}
        \mbf{x}_{k-1\mid D} &= \mbf{x}_{k-1\mid k-1} + \mbf{c}_{k-1}\big(\mbf{x}_{k\mid D} - \mbf{x}_{k\mid k-1}\big),
        \\
        \mbf{P}_{k-1\mid D} &= \mbf{P}_{k-1\mid k-1} +\mbf{c}_{k-1}\big(\mbf{P}_{k\mid D} - \mbf{P}_{k\mid k-1}\big)\mbf{c}_{k-1}^T,
     \end{align}
    where  the matrix $\mbf{c}_{k-1}$ back-propagates the information,
    \begin{equation}
        \mbf{c}_{k-1} = \mbf{P}_{k-1\mid k-1}\mbf{F}^T\mbf{P}_{k\mid k-1}^{-1}.
    \end{equation}

    RTS smoothing is a powerful technique, in no small part because it helps
    correct for mistakes made when running the Kalman filter. An obvious
    correction is in the eponymous smoothing of the stream track. Also important
    is that the smoother will update the initial conditions $(\mbf{x}_0$,
    $\mbf{P}_0$), which are only motivated by a heuristic -- a local average at
    the starting point. In particular the covariance $\mbf{P}_{0\mid 0}$ should
    have off-diagonal terms as the measurement error is correlated in the real
    data. Determining good values of these initial off-diagonal terms is neither
    easy nor necessary as the RTS smoothing back-propagates information on the
    covariance from later time steps to improve $\mbf{P}_0$.

\begin{figure}
      \script{method_kalman-filter.py}
      \centering
      \includegraphics[width=1.0\linewidth]{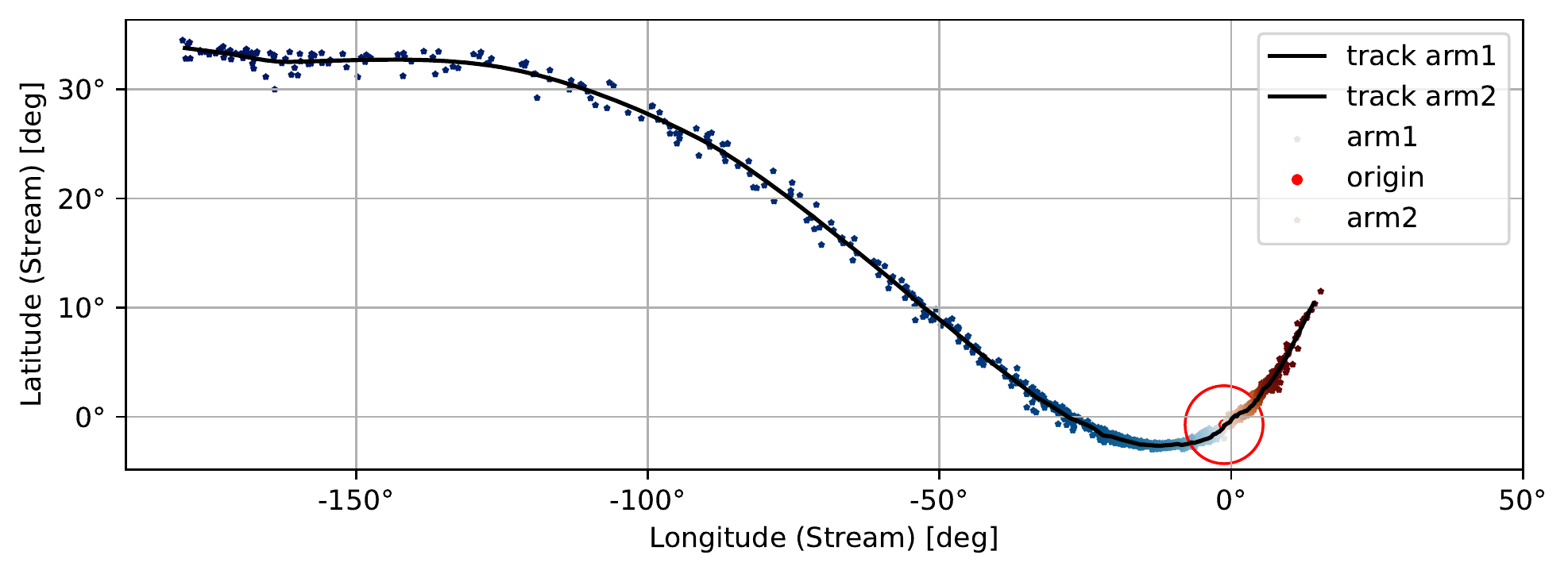}
      \caption{ \emph{The Kalman-predicted path for \stream{NGC-5466}}. Building
        on figures
        \ref{fig:method_frame-fit},\ref{fig:method_som-fit},\ref{fig:method_time-proxy},
        this figure shows the application of the Kalman filter to each arm of
        the simulated \stream{NGC-5466}. The ordering of the data in each arm
        was done by SOM and the data re-interpreted as a time-series by using
        the point-to-point distance in the projection onto the SOM 1-D subspace.
        The Kalman filter was run on this data to produce the stream tracks
        (black). }
      \label{fig:method_kalman-filter}
    \end{figure}

    In \autoref{fig:method_kalman-filter} the Kalman filter pipeline is run on
    the SOM-ordered data from the mock NGC-5466 stream shown in prior figures.
    This Kalman filter was run in spherical coordinates, with the corresponding
    distance metric. The stream track correctly accounts for the angle wrapping
    at $-180^\circ$ to $+180^\circ$. The track covariances (gray ellipses at
    time steps $\Delta t_k$) increase where the data are dense and the stream is
    cold and decreases elsewhere, such as as at the end-points of the stream.
    For most of the stream the covariance ellipses are the given width of the
    stream, $2^\circ$, since the data are relatively dense and the errors in
    on-sky coordinates are 0 in this mock stream and negligible in, e.g. \gaia{}
    data, so do not contribute to the stream width as determined by the Kalman
    filter.

\section{Results} \label{sec:results}

  Our method of fitting a track to a stream happens in three steps. First
  (\autoref{sub:great_circle_coordinate_frame}) we find a reference frame to
  optimally represent the data. Second (\autoref{sub:self_organizing_maps}) we
  find a one-dimensional subspace which approximately matches the structure of
  the stream and into which the stream can be projected. Third
  (\autoref{sec:kalman_filter}), having ordered the data, we fit the track by
  Kalman filter. This three-step method produces good fits to the example data
  sets used thus far. In this section we show the application of the method
  pipeline to a variety of stellar streams, simulated using a number of
  different techniques.

  \subsection{Comparing to Distribution Function (DF) models}
  \label{sub:comparing_to_distribution_function_df_models}

\begin{figure*}
      \script{results_streamdf.py}
      \centering
      \includegraphics[width=0.9\linewidth]{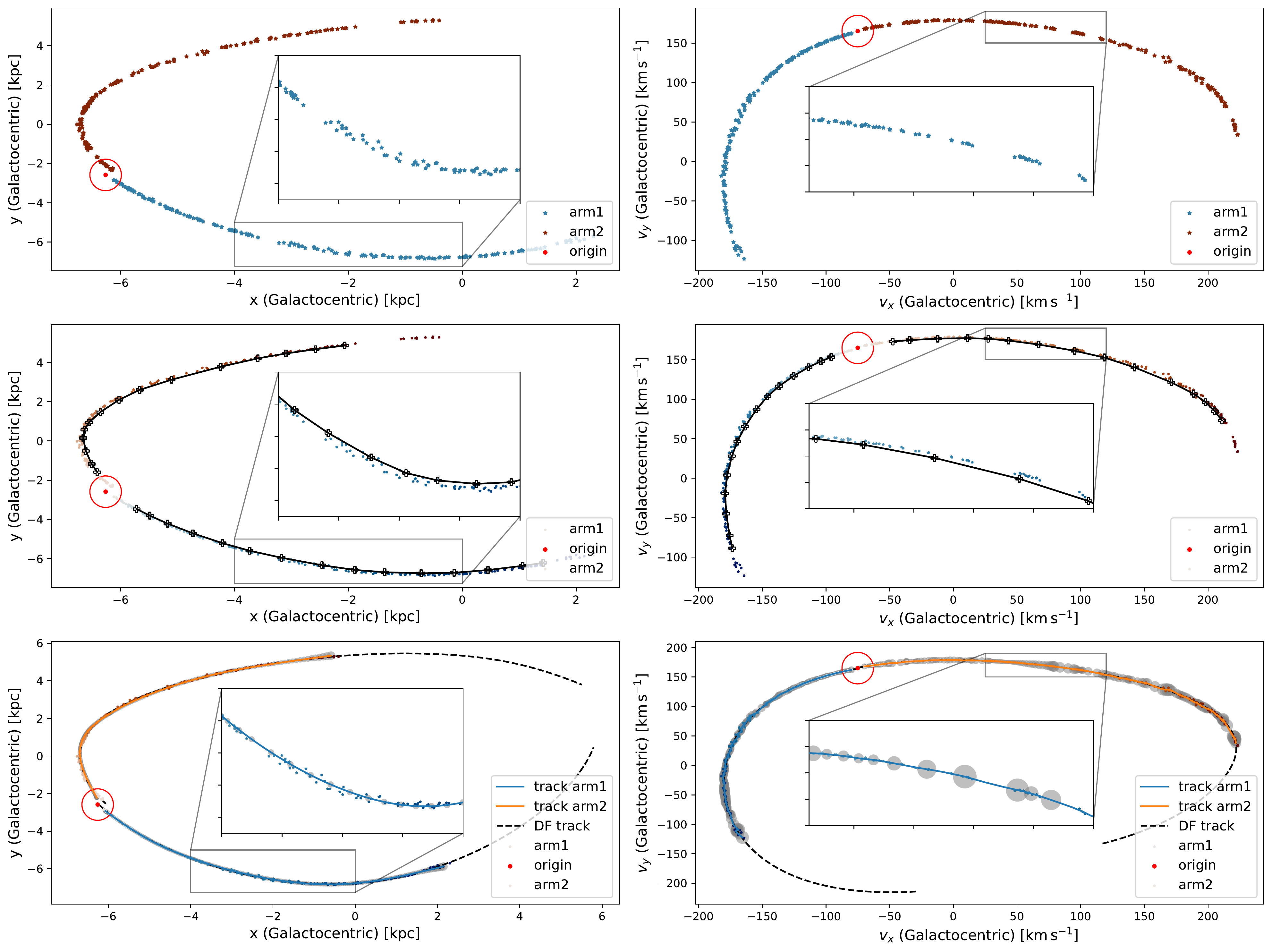}
      \caption{ 
        \emph{Comparing results of \galpystreamdf{}} \citep{Bovy2014} to
        \trackstream{}. The stream progenitor parameters are \stream{NGC 104}
        from \citet{Vasiliev2019}, integrated in \galpyMWPotential{}
        \citep{Bovy2015}, with \galpystreamdf{} from \citet{Bovy2014}.
        [top left/right] x-y and $v_x$-$v_y$ projections of the simulated
        stream's arms and progenitor. The data are not yet ordered, beyond
        identifying to which stream arm each tail belongs.
        [middle left/right] Each stream arm is ordered by a SOM
        (\autoref{sub:self_organizing_maps}) run on the positions and kinematics
        of each stream-arm. The trained SOM prototypes are black pluses.
        [bottom left/right] The \trackstream{} tracks (orange and blue, with
        gray error ellipses) are compared to the \galpystreamdf's true tracks
        (dashed line), also the source of the 400 star sample on which the
        \trackstream{} was trained. The error ellipses, shown at each step of
        the Kalman filter, are smoothly interpolable as a function of the
        arc-length. For visualization of the covariances we exaggerate the
        intrinsic stream width to 300 pc. See
        \autoref{sub:comparing_to_distribution_function_df_models} for
        discussion.
        \label{fig:results_streamdf}
      }
    \end{figure*}

    In \citet{Bovy2014} stellar streams are modeled with a distribution function
    in action angle space. Starting from a potential, a progenitor position, and
    stream formation parameterization at some initial time $t_0$, the initial
    stream distribution is evolved to time $t$. In action-angle space the
    progenitor and stream is represented very simply. The action-angle stream
    distribution is transformed to position-momentum phase space, and similarly
    parameterized using linear interpolations. The DF method simulates an exact
    stream track and width. \autoref{fig:results_streamdf} shows a mock stream
    of \stream{NGC 104}, simulated with \galpystreamdf{} from \citet{Bovy2014}.
    The progenitor's current phase-space position is from \citet{Vasiliev2019}
    with a total disruption time 4.5 Gyr and an initial velocity dispersion of
    0.365 km / s, to create a reasonably thin stream. The system is integrated
    in \galpyMWPotential{} with an
    \galpyapidoc{reference/aaisochroneapprox}{actionAngleIsochroneApprox} to map
    phase space coordinates to the action-angle coordinates required by the
    mock-stream generator \citep{Bovy2015}.

    In contrast the \trackstream{} pipeline estimates the track from stream
    sample points (e.g. stars) in position-momentum phase space. To compare
    \trackstream{} to a DF prediction we sample from the stream, transforming
    from action-angles space, and run \trackstream{} on the sample.
    \autoref{fig:results_streamdf} shows the result and comparison to the
    DF-derived stream path. For visualization we use an intrinsic stream width
    ($\sigma_{\rm{stream}}$) of 300 pc, which exaggerates the size of the
    covariance ellipses. When run with the more accurate intrinsic stream width
    of 100 pc the two methods agree within $1-\sigma$ over their lengths. The DF
    track extends beyond that of \trackstream{} since there are no tracer
    particles sampled at those distances. Another major difference between the
    two methods is what is meant by the track width. The DF analytically
    calculates the intrinsic width of the stream, which is a 1-$\sigma$ Gaussian
    along the track. \trackstream{} combines an intrinsic stream width with the
    uncertainty in each datum point and the sparsity of the data to derive a
    covariance of the confidence in the track. This difference can be seen in
    \autoref{fig:results_streamdf}, especially in regions of data sparsity, like
    the poorly-sampled outer lengths of the stream. Here the confidence in the
    stream track is quite low. For a large number of well-measured data points
    \trackstream{} and the DF converge, such as near the progenitor in
    \autoref{fig:results_streamdf}.

    The stream DF approach of \citet{Bovy2014} and similar (e.g.
    \citet{Sanders:2014}) are some of the few analytic means to simulate stellar
    streams. However the method is subject to a number of limitations. First, it
    is generally necessary in the progenitor model to model the stripping times
    of stars from the progenitor to a uniform distribution. This assumption is
    known to be false, since the mass loss rate is highest near pericenter
    passages. Second, these methods do not account for the progenitor's
    gravitational influence on the stream. The next stream-generation method
    trades analyticity for the capability to address many of these limitations.


  \subsection{Tracking particle spray mock streams}
  \label{sub:tracking_particle_spray_mock_streams}

    Particle-spray (also called release) methods are a semi-analytic model of
    stream formation. The model requires a potential, the current phase-space
    position of the progenitor, the mass of the progenitor, and the period of
    time and rate at which the progenitor has been disrupting. To model the
    internal dynamics, the model also often includes a parameter for the
    distribution of speeds at which particles are ejected.

    First the phase-space position of the progenitor is backwards integrated to
    the start of the disruption period. If the model accounts for mass loss, the
    progenitor's current mass may also be back-integrated to the initial mass by
    considering the disruption rate. Next, the progenitor is forward integrated
    to the present, periodically releasing particles (stars) according to the
    disruption prescription. The particles are released on orbits from a
    simulation-calibrated distribution of initial positions/velocities of
    released particles. When fully integrated the progenitor is at its present
    position. Despite the name, all released particles are stream stars and
    should lie along the stream track, with a reasonable dispersion.

\begin{figure*}
      \script{results_streamspray.py}
      \centering
      \includegraphics[width=0.9\linewidth]{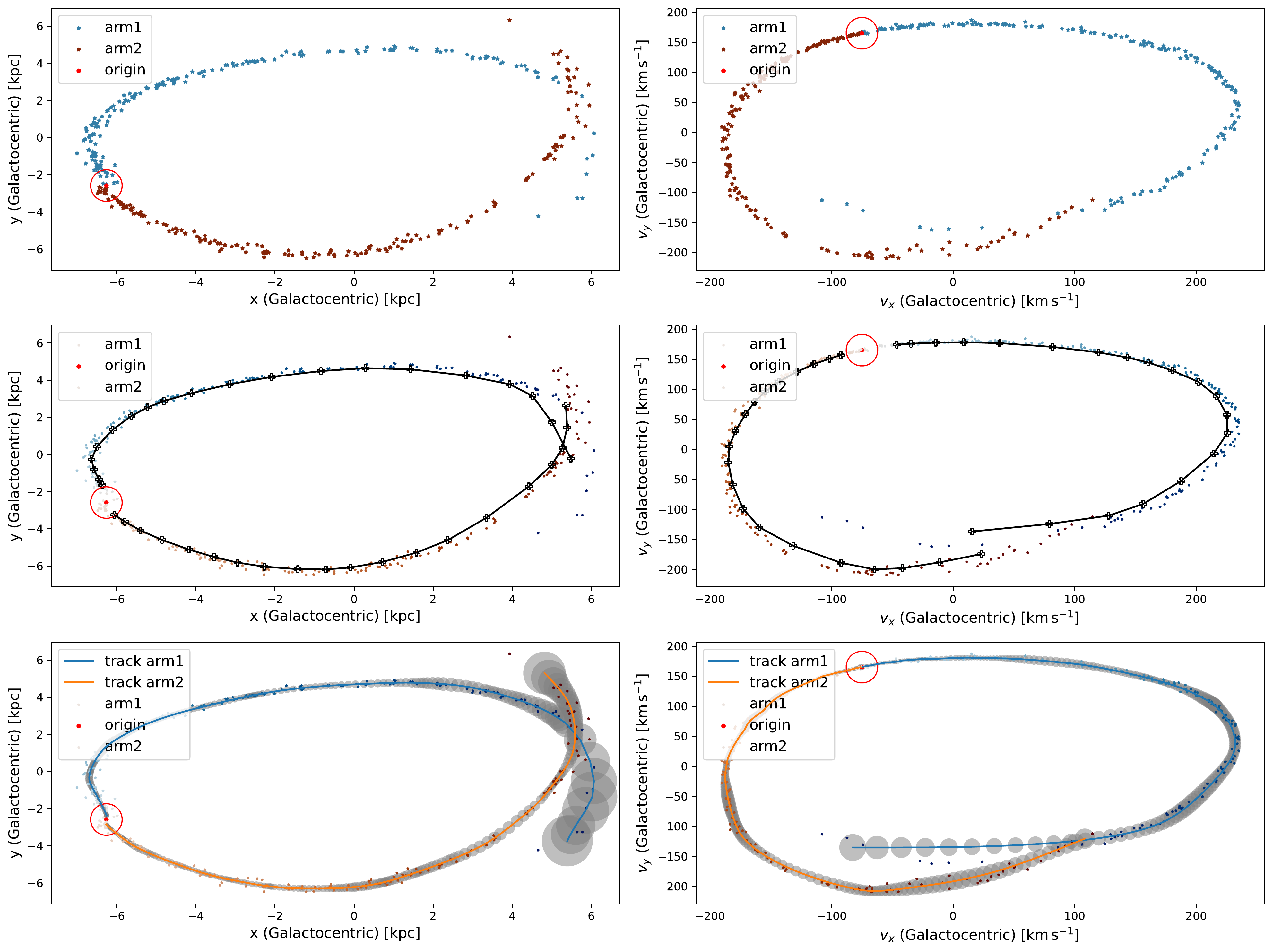}
      \caption{
        \emph{Full application of \trackstream{} to a particle-spray model of
        \stream{NGC 104}.} The mock stream progenitor parameters are \stream{NGC
        104} from \citet{Vasiliev2019}, integrated in \galpyMWPotential{}, with
        \galpystreamspraydf{} from \citep{Bovy2014}.
        [top left/right] x-y and $v_x$-$v_y$ projections of the simulated
        stream's arms and progenitor. The data are not yet ordered, beyond
        identifying to which stream arm each tail belongs.
        [middle left/right] The stream is ordered (and colored) by the SOM
        \autoref{sub:self_organizing_maps} run on the positions of each arm. The
        trained SOM prototypes are black, connected pluses.
        [bottom left/right] The \trackstream{} track is compared to
        \galpystreamspraydf{} track, the source of of the 400 star sample on
        which the \trackstream{} was trained. For visualization of the
        covariances we exaggerate the intrinsic stream width to 300 pc.See
        \autoref{sub:tracking_particle_spray_mock_streams} for discussion. }
      \label{fig:results_streamspray}
    \end{figure*}

    \autoref{fig:results_streamspray} shows a particle-spray generated mock
    stream with a 47 Tucanae-like progenitor. We use the particle-spray
    implementation -- \galpystreamspraydf{} -- that is included in \galpy{}
    (v1.8+) \citep{Bovy2015} , which is an implementation of the
    \citet{Fardal2015} method that is fully described in \citet{Qian2022}. The
    potential is \galpyMWPotential{} from \citet{Bovy2014}; the progenitor's
    orbital parameters are from \citet{Vasiliev2019} with the mass from
    \citet{MarksKroupa2010} (We note that \galpystreamspraydf{} does not
    incorporate mass loss, so this mass is fixed). The disruption time is 2 Gyr,
    chosen arbitrarily to generate a sufficiently long mock stream. All other
    parameters are set to the \galpystreamspraydf{} defaults. The stream track
    is fit with the \trackstream{} pipeline. As in
    \autoref{sub:comparing_to_distribution_function_df_models}, for
    visualization purposes we set the intrinsic stream width to 300 pc.
    Particle-spray methods do not model the stream track, so the fit quality is
    seen only by comparing to the tracers. The tracks for both arms correctly
    start at an offset from the progenitor. Though the initial state of the
    Kalman filter is much closer to the progenitor, the RTS smoother correctly
    back-propagates the information that the stream does not start at the
    progenitor. As expected, the uncertainty in the track is larger at the start
    than immediately after. While the stream tracks are smooth, they exhibit
    some structure at small scales due to variations in the distribution of the
    stream stars. The structure in the track is a balance between the
    information known about the stream -- e.g the intrinsic width -- and the
    actual structure of the data. If the data has sufficient structure, be it
    random from low-number statistics or from e.g. subhalo interactions, so too
    will the \trackstream{} stream track. It is possible to damp structure in
    the track by including smoothness priors. If any sources of stream
    perturbations are included in the simulation, such as sub-halos, then these
    priors would likely be quite complex. Alternatively, in simulations one can
    generate more tracer particles, providing more information when constructing
    the stream.

  \subsection{Tracking an \nbody{} model}
  \label{sub:tracking_an_n_body_model}

    \nbodys{} are the most accurate but also computationally expensive of the
    stream simulation methods, making the fewest time-saving approximations. A
    high resolution \nbody{} simulation may achieve one-to-one parity in the
    number of particles to number of stars in the progenitor, model the motions
    with dynamic time steps to resolve close interactions, and include stellar
    evolution. The disruption rate is determined by the the Galactic tidal field
    and the progenitor's internal kinematics, not an analytic prescription.
    Often the high resolution of the progenitor is not matched in the potential,
    where it is common to instead use a purely analytic model or to also include
    low(er) resolution sub-halos. When trying to match observational data, an
    \nbody{} simulation provides the most accurate representation.

\begin{figure*}
      \script{results_nbody.py}
      \centering
      \includegraphics[width=0.9\linewidth]{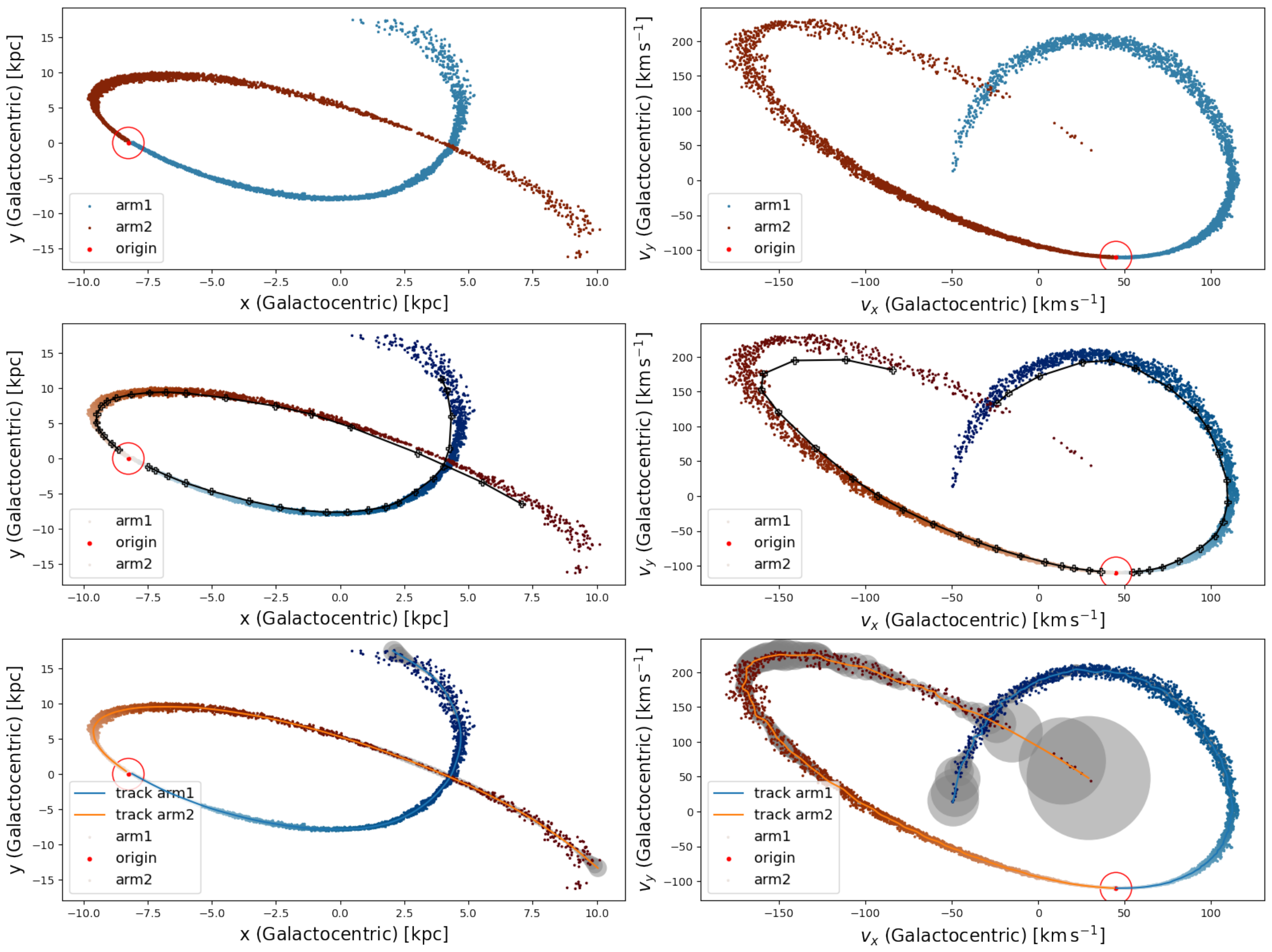}
      \caption{
        \emph{Full application of \trackstream{} to an \nbody{} simulation of
        \stream{Palomar 5}.} The mock stream progenitor parameters are for
        \stream{Palomar 5}, integrated in \galpyMWPotential.
        [top left/right] x-y and $v_x$-$v_y$ projections of the simulated
        stream's arms and progenitor. The data are not yet ordered, beyond
        identifying to which stream arm each tail belongs.
        [middle left/right] The stream is ordered by the SOM
        \autoref{sub:self_organizing_maps} run on the positions of each arm. The
        trained SOM prototypes are black, connected pluses.
        [bottom left/right] The \trackstream{} track is compared to the
        \nbody{}, the data on which the \trackstream{} was trained. For
        visualization of the covariances we exaggerate the intrinsic stream
        width to 300 pc. See
        \autoref{sub:comparing_to_distribution_function_df_models} for details.
        }
      \label{fig:results_nbody}
    \end{figure*}

    \autoref{fig:results_nbody} shows \trackstream{} applied to an \nbody{} of
    \stream{Palomar 5}. The \nbody{} is taken from \citet{Starkman2019}, where
    we searched for extended tidal features of \stream{Palomar 5}. Further
    details of the simulations can be found in \citet{Starkman2019}. In short,
    we use the direct $N$-body code \nbodysix{} \citep{Aarseth2006}, evolving an
    initial progenitor of a 100,000 star Plummer model with a half-mass radius
    of 10 pc. The initial mass function is from \citep{Kroupa2001ki}, with a
    minimum and maximum stellar mass of $0.1 M_\odot$ and $50 M_\odot$,
    respectively. We use the stellar evolution prescription of
    \citet{Hurley2000}, with a metallicity of $Z=0.001$, and \citet{Hurley2002}
    for binary stars. The cluster is evolved for 12 Gyr in \galpyMWPotential
    such that the final position, velocity, mass, and mass function are
    comparable to the observed properties of \stream{Palomar 5}
    \citep{Ibata2017, Grillmair2001}, though the simulated cluster density is
    higher than observed.

    The progenitor particles are removed as these are not part of the stream.
    Likewise we remove the neutron stars and other stellar remnants. The full
    \nbody{} simulation includes a number of particles that were ejected from
    the progenitor, e.g. in a close core interaction, that are not part of the
    stream. These non-stream members are identified by a Local Outlier Factor
    (LOF) calculation \citep{Breunig2000} with a threshold of $1.0$ and removed.
    Approximately, the LOF is a local density calculation. Since the simulated
    stream is dense, low density regions are likely not in the stream. The
    distribution of LOFs calculated for each star is very strongly peaked around
    the typical density of the stream. Only in the extended tidal tails is the
    LOF of the stream approximately that of the non-stream members. In
    observations it is unlikely these stars would have a high stream membership
    probability. The exact value of the LOF threshold often makes little
    difference and we find the stream track determination robust.

    After removing outliers, \trackstream{} is applied to the data. Starting
    near the progenitor, like for
    \autoref{sub:tracking_particle_spray_mock_streams}, the tracks for both arms
    are offset from the progenitor center. At each point along the stream, the
    RTS-smoothed Kalman filter uses information both proceeding and preceding
    that point to construct the track, which smoothly fits the stream data. In
    higher density, kinematically cold regions the width (covariance ellipses)
    shrink down to the prescribed stream width, which is exaggerated here to 300
    pc for visualization purposes. As expected, where the stream density
    decreases or the stream is kinematically hotter the track confidence
    decreases and the covariance ellipses grow. At the extrema of each arm's
    track the uncertainty is also large since there are fewer stars and because
    at extrema the Kalman filter only has information on one side, not both. To
    the Kalman filter, the start of the stream by the progenitor is also an
    extremum. Physically, the progenitor is not an extrema, since the other
    stream arm in \autoref{fig:results_nbody} is nearby and shares the
    progenitor as the origin. The Kalman filter could be modified to account for
    the presence of other stream arm, but this is probably not advisable.
    Theoretically, fitting the Kalman filter to each stream arm could be done as
    an ensemble fit, with the progenitor as a shared constraint. While this
    modification would reduce the extrema-associated uncertainty, it is unclear
    how to perform this ensemble fit without assuming information about the
    potential and evolution of the cluster, assumptions we purposefully avoid.
    Regardless, near the progenitor the star count is usually larger than
    elsewhere along the stream and the track certainty is correspondingly
    higher, mitigating the need for an ensemble fit.

  \subsection{Tracking Real Streams}
  \label{sub:tracking_real_streams}

\begin{figure}
      \script{results_Pal5_real.py}
      \centering
      \includegraphics[width=0.9\linewidth]{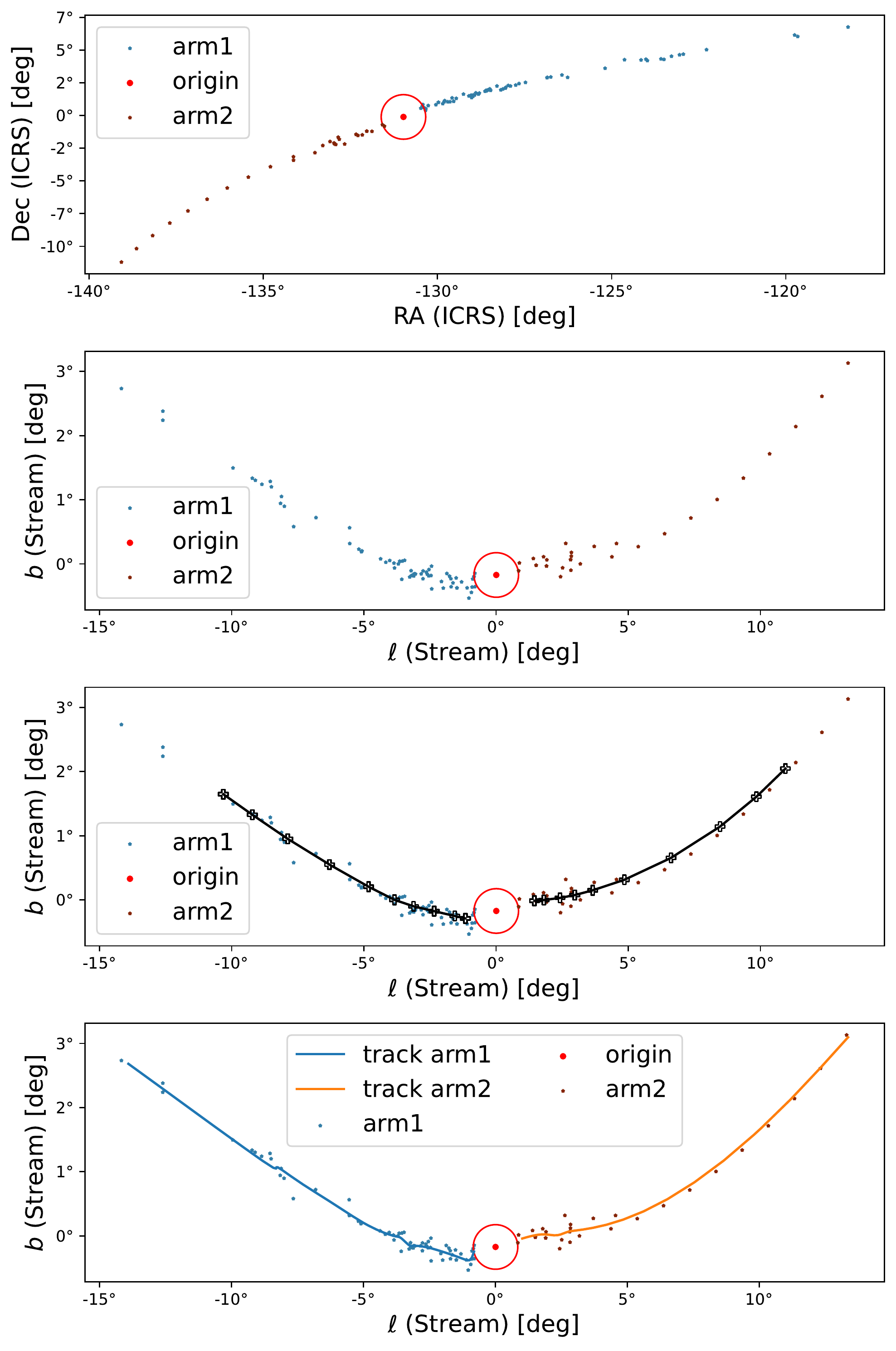}
      \caption{\label{fig:results_Pal5_real}
        \textit{Full application of \trackstream{} to real data on the
        \stream{Palomar 5} stream from \citet{Ibata2017}
        and mean track points from \citet{Starkman2019}}.
        [top] The stream stream is shown in ICRS (observed) coordinates. The
        position of the origin is from \citet{Vasiliev2019}. We hand-label and
        color the stream arms.
        [top middle] A best-oriented frame is fit to the data, minimizing $b$.
        [bottom middle] An SOM (black connected dots) is fit to each arm. 
        [bottom] The Kalman filter is run on each arm (blue and orange), using
        the SOM ordering.
    }
    \end{figure}

\begin{figure}
      \script{results_GD1_real.py}
      \centering
      \includegraphics[width=0.9\linewidth]{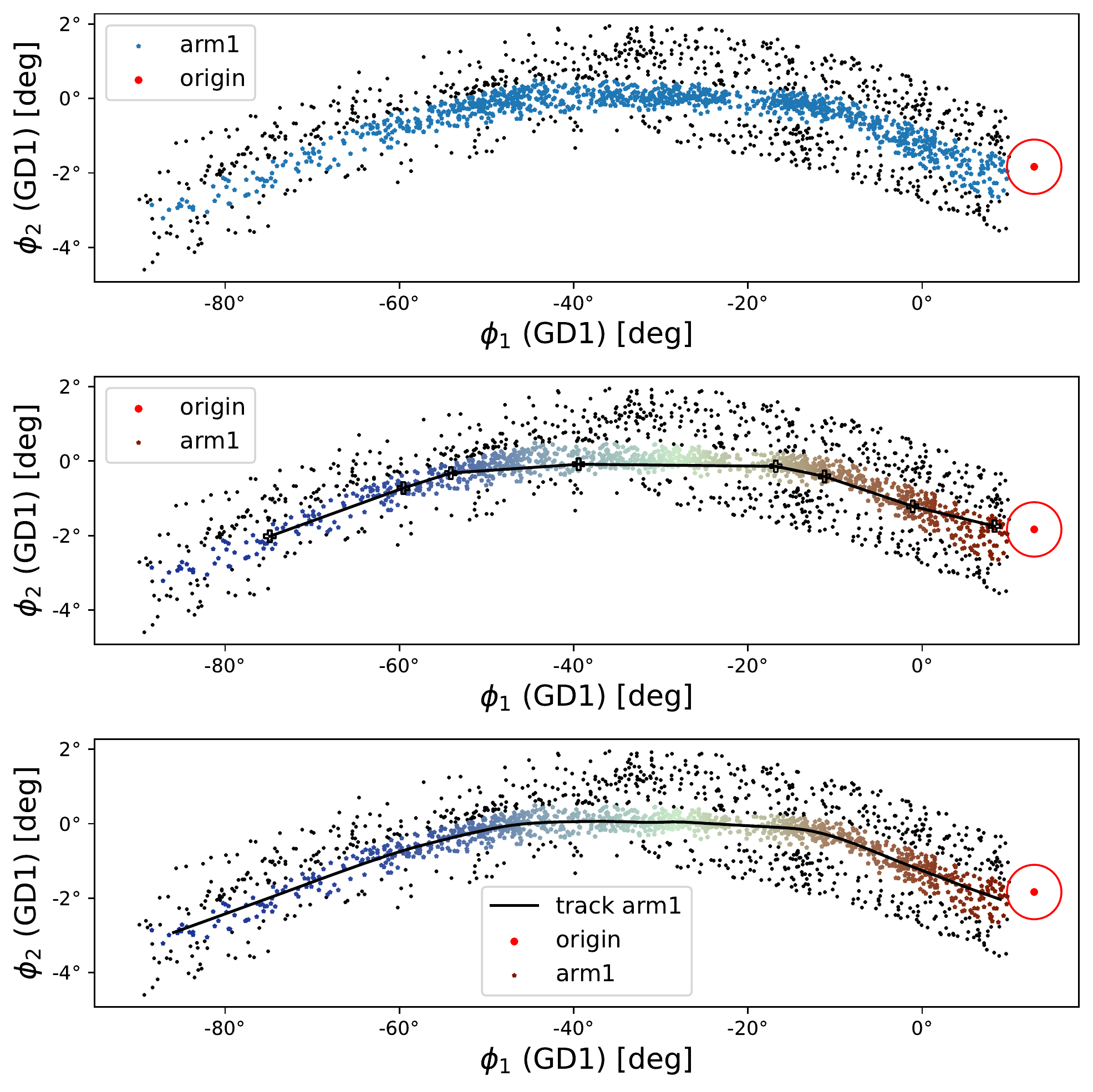}
      \caption{
        \textit{\trackstream{} fit to a subset of the \stream{GD-1} data from
        \citet{Price-Whelan2018}}.
        [top] The data transformed to the frame from \citet{Koposov2010}. The
        thin stream is hand-selected in TOPCAT \citep{Taylor2005} to match the
        selection field of \citet[][figure 2]{Price-Whelan2018}.
        [middle] The same, but now including the SOM prototypes (black connected
        points) and coloring the data by the SOM ordering.
        [bottom] The fit track to the stream (black) overlaid on the SOM-colored
      data. }
      \label{fig:results_GD1_real}
    \end{figure}

    In the previous subsections of \autoref{sec:results}, \trackstream{} was
    applied to mock stream data. \trackstream{} also works on real data. We
    apply the pipeline to two of the most commonly studied stellar streams:
    Palomar 5 and GD-1, shown in \autoref{fig:results_Pal5_real} and
    \autoref{fig:results_GD1_real}, respectively. The Palomar 5 data are a
    combination of data from \citet{Ibata2017} and \citet{Starkman2019}. In the
    stream-oriented reference frame the stream stars span almost 30 degrees in
    longitude but only a few degrees in latitude. The data set is very small,
    demonstrating that \trackstream{} fits reasonable tracks even for
    $\mcal{O}(10)$ stars. Like with prior examples, the confidence in the track
    is a mixture of the intrinsic stream width, the errors in the data, and the
    data sparsity. Near the ends of the track, particularly the end point at
    negative longitude, the covariance ellipse is much larger than elsewhere in
    the stream. Looking at the nearby data, two reasons for the large covariance
    are that the data are sparse and that the end regions have less information
    (a hard limit to on one side) to inform the fit. Were the data less sparse
    the covariance would be more similar to the covariance at the other end
    point. \autoref{fig:results_GD1_real} shows a much larger, though still
    small compared to e.g. the \nbody, data set from \citet{Price-Whelan2018} of
    \stream{GD-1}. The thin stream is hand-selected in TOPCAT \citep{Taylor2005}
    to match the selection field of \citet[][figure 2]{Price-Whelan2018}. The
    progenitor of GD-1 is unknown, so for convenience we treat the whole stream
    as a single stream arm, rather than a leading and trailing arm.

\section{Discussion and Conclusions} \label{sec:discussion}

    In this paper we present a novel method for rapidly constructing stellar
    stream tracks, accounting for measurement error and data sparsity  (e.g.
    Fig. \ref{fig:results_streamspray}). By using stream-oriented
    reference-frame transformations in conjunction with Self-Organizing Maps, we
    treat stellar-stream data as a pseudo time-series, to which first-order
    Kalman filters can be applied. The technique is Galactic-model independent,
    non-parametric, and works on many different mock stream generation methods.
    Owing to its speed this method is well suited to comparing simulations to
    data as a step in a Monte Carlo analysis.

    The track-fitting procedure, implemented in
    \href{https://github.com/nstarman/trackstream}{\trackstream{}}, is two or
    three steps, depending on if one is working in spherical coordinates. First,
    if in spherical coordinates we re-orient the reference frame to minimize the
    training time of the following step. Second, using SOMs we infer a
    one-dimensional subspace that approximates the structure of the target
    stream and orders the constituent data. Third and last, using the
    subspace-order we fit the track of the stream with a Kalman filter. The
    track-finding technique works well for a variety of stream-generation
    methods: distribution functions
    (\autoref{sub:comparing_to_distribution_function_df_models}), particle
    sprays (\autoref{sub:tracking_particle_spray_mock_streams}), \nbodys{}
    (\autoref{sub:tracking_an_n_body_model}), even real streams
    (\autoref{sub:tracking_real_streams}).

    In simulations, outliers can be present. In nature, contamination by
    background or foreground stars can also occur. \trackstream{} assumes that
    all stars are members of the stream and should therefore contribute to the
    track fit: the SOM will include the outliers and contamination when finding
    the stream's subspace and the Kalman filter will likewise try to move the
    mean path towards those points. Without a background model, contamination is
    challenging to deal with; however if stream membership probabilities are
    known \emph{a-priori}, than the SOM and Kalman filter can both incorporate a
    weighting (the probability) to minimize the impact of low-probability
    members, which are likely contaminants. Outliers may be detected and
    removed, in separate steps to the SOM and Kalman filter. In the case of
    \autoref{fig:results_nbody}, a LOF metric is used to rank for each datum how
    connected it is to neighboring data. Data with connectivity below a given
    threshold are marked as an outlier and removed. This initial step of outlier
    removal is very useful and sufficient in all the examples discussed here.
    For more challenging scenarios we find it useful to add a second outlier
    removal step, after running the SOM. Now the LOF is run on the data
    projected into the SOM manifold, which strongly constrains what data are (or
    are not) neighbors. The SOM may be re-run, or just further train, to
    minimize the impact of the outliers on the SOM manifold. Alternatively, one
    could modify the Kalman filter to use Gaussian distribution conditioned on
    an adaptive width, essentially a hierarchical model where the width has its
    own model. This modified Kalman filter will be less sensitive to outliers.
    In this paper we employ only the single outlier detection and removal step,
    which is sufficient for these simulated and real streams.

    One of the motivating use cases of \trackstream{} is to fit stream tracks
    with no human supervision, for example as a step in a likelihood calculation
    of observational data given the stream model in an MCMC. \trackstream{} has
    a number of inputs, some requiring a human check to ensure the track is well
    fit. In particular, the SOM needs to be checked for convergence, which can
    require a long training period if the initial prototypes are poorly chosen.
    Since human intervention is not possible at every step in an MCMC,
    \trackstream{} is designed such that the inputs only need to be well chosen
    once and then apply well for any stream generated in a similar region of
    parameter space. For inputs like the initial stream width, these are
    generally well known and a stream with a very different width might produce
    a less smooth track, but covariance ellipses will likewise be large and the
    model likelihood will be low. For the SOM, using the initial prototypes also
    means having a long training time at every step. We emphasize that ``long''
    is of order 10 seconds for $10^5$ iterations on a consumer laptop, but this
    may still be a prohibitive time constraint. Steps in the MCMC are to nearby
    points in the parameter space of a potential and progenitor and the
    generated mock stream will generally be similar to the previous stream. For
    subsequent steps, the previous SOM may be used as an informed starting point
    and should require only a little tuning, not wholesale retraining, and most
    importantly, no human supervision.

    For reference we compare the training time of a SOM to the DF method
    \citep{Bovy2014}, which is the fastest of the simulation methods described
    here. The speed of the DF method depends on the underlying potential: with a
    simple logarithmic potential both arms of a stream with a \stream{NGC
    5466}-like progenitor may be simulated in approximately 20 seconds on a 2018
    2.7GHz i7 MacBook Pro, while a similar progenitor evolved in
    \galpyMWPotential{} takes approximately $2\frac{1}{2}$ minutes. Training the
    Self-Organizing Map for 2 million iterations takes approximately 20 seconds
    (ten per arm) on either stream. For well chosen initial prototype locations
    the number of iterations may be reduced arbitrarily, but running for at
    least a few tens of thousands of iterations can be good practice to ensure
    convergence.

    The stream is re-simulated with a modified \galpyMWPotential, where the mass
    of the bulge and disc are both increased by 30\% -- a very large difference
    and one that significantly affects the stream morphology, with a very
    obvious kink in the z projection. Retraining from scratch would take a few
    million iterations and tens of seconds. However, the SOM prototypes trained
    on the stream simulated with \galpyMWPotential{} are proximate in
    phase-space and already have the correct topological arrangement. Using
    these prototypes as an informed initial condition, the requisite training
    time to convergence is reduced to a few tens of thousands of iterations and
    less than one second per arm. The SOM is only a time-limiting step of the
    \trackstream{} pipeline once. The Kalman filter is never a time-limiting
    step, taking only a few milliseconds to run. Thereafter, the entire pipeline
    is significantly faster than the stream's generative model.

    \trackstream{}, in its current form, computes an affine parameterized mean
    track and its associated covariance. The latter incorporates the intrinsic
    width, but it is hard to separate this from the details of the covariance. A
    useful future improvement to \trackstream{} will be to modify the Kalman
    filter to be a Bayesian hierarchical model incorporating the intrinsic
    width, which would be constrained simultaneously with the mean path. In
    addition, the Kalman filter can be relaxed to the more general Bayesian
    filter, allowing for non-Gaussian descriptions of the streams.

    \trackstream{} is an efficient means to characterize a stream's path. While
    designed for use with simulations, \trackstream{}'s methods and pipeline
    work well with real observational data. For example, a probabilistic model
    can use the stream track to compute the likelihood of a mock stream given
    real data, or vice versa, the probability of the data given a mock stream.
    Running \trackstream{} on both the observations and simulations allows for
    the comparison of both tracks. Therefore, in application, \trackstream's
    novel method for rapidly constructing stellar stream tracks enables us to
    approach a problem using the path of real data, simulated data, or both.



\section*{Acknowledgements} \label{sec:acknowledgements}

  We thank Joshua Speagle for insight and means to incorporate data error into
  SOM weighting. NS acknowledges support from the Natural Sciences and
  Engineering Research Council of Canada (NSERC) - Canadian Graduate
  Scholarships Doctorate Program [funding reference number 547219 - 2020]. NS
  and JB  received partial support from NSERC (funding reference number
  RGPIN-2020-04712) and from an Ontario Early Researcher Award (ER16-12-061; PI
  Bovy).



\section*{Data Availability} \label{sec:data_availability}

  Using ShowYourWork \citep{Luger2021}, all the figures in this paper can be
  reproduced from their source code at
  \href{https://github.com/nstarman/trackstream_paper}{https://github.com/nstarman/trackstream\_paper}.
  \trackstream{} may be used under a modified BSD-3 license and is available
  on GitHub in the repository
  \href{https://github.com/nstarman/trackstream}{https://github.com/nstarman/trackstream}.



\bibliographystyle{mnras}
\bibliography{bib}


\appendix
\onecolumn

\section{The Math for a 1-D SOM}
\label{app:som_math}

    The goal of SOM is to sequentially introduce data vectors $\mbf{w}_n$ and
    adapt the prototypes and lattice according to the data.

    See \autoref{som_algorithm_steps}.

    \subsection*{\tbf{(1) Given the data in the stream coordinates:}}

      \begin{align}
        \intertext{Let the data be a finite number $\mathrm{N}$ of vectors, indexed by $n$, with $\mathrm{D}$ features (like 2 or 3 positions and 2 or 3 velocities).}
          \mbf{w}^{(n)} & \in \mathds{R}^\mathrm{D} & \text{where $n \in \mathds{N} \cap [1, \mathrm{N}]$.}
        \intertext{In matrix form,}
          \mbf{w} &= \left[\mbf{w}^{(1)} \mathellipsis \ \mbf{w}^{(N)}\right]^T \in \mathds{R}^{\mathrm{N} \times \mathrm{D}}.
      \end{align}

    \subsection*{\tbf{(2) Choose $\mbf{K}$ prototypes $\mbf{p}^{(k)}$ with
    linear lattice $\mbf{Q}$:}}

      \begin{align}
        \intertext{The SOM learns the organization -- i.e. the 1-D structure -- of the data with $K$ prototype vectors in the data's space}
          \mbf{p}_k &\in \mathds{R}^\mathrm{D}, & k \in \mathds{N} \cap [1, \mathrm{K}].
        \intertext{In matrix form,}
          \mbf{p} &= \left[\mbf{p}^{(1)} \mathellipsis \ \mbf{p}^{(K)}\right]^T \in \mathds{R}^{\mathrm{K} \times \mathrm{D}}.
        \intertext{
          The prototype vector topology is tracked with a feature map $\mbf{Q}$,
          where each prototype has a corresponding lattice point $\mbf{q}_{k}$.
          A one-dimensional SOM means $\mbf{Q} \in {\mathds{Z}^+}^{\mathrm{K}} $ is one-dimensional
          and the only information contained in $\mbf{q}_{k}$ is its index:
        }
          \mbf{q}_k &\equiv k.
        \intertext{
          This is called a linear lattice. With a linear lattice, prototype vectors $p_i$, $p_j$ are neighbors
          if and only if the corresponding lattice points $q_i$, $q_j$ are
          neighbors -- i.e. $|i-j| = 1$. The distance between any two prototypes is given by the distance matrix $\mbf{D}_Q\in\mathds{R}^{K \times K}$, where
        }
          d_{ij} &= \norm{\mbf{p}^{(i)} - \mbf{p}^{(j)}}_Q = \norm{q_i - q_j}\equiv | i - j |,  \label{eq:app:som_lattice_distance}
      \end{align}

      The prototypes can be randomly initialized or more intelligently chosen.
      The goal of the next step is to train the SOM on the data.

    \subsection*{\tbf{(3) Iteratively:}}

      \subsubsection*{\tbf{(1) Select the next datum $\mbf{w}^{(n)}$}}

        The data are selected until exhaustion and without replacement to the
        SOM. The SOM is generally run on many such selection cycles until it
        reaches equilibrium. While the ordering of the data should not be
        important, the SOM is not guaranteed to converge to the global minimum,
        so in principle the order can be consequential. However, for all
        examples presented in this paper, the ordering has not proved important
        and is not further considered.

      \subsubsection*{\tbf{(2,3) Update the position of the nearest prototype
      and its neighbors}}

        Having selected a data point, the prototypes need to be updated in both
        the data and lattice spaces -- we start with the closest.

        Using whatever distance metric is appropriate for the data space, the
        nearest prototype is $\mbf{p}^{(c(n))}$, i.e. such that
        \begin{equation} \label{eq:app:som_BMU}
            \norm{\mbf{p}^{(c(n))} - \mbf{w}^{(n)}} = \rm{min}\{\norm{\mbf{p}^{(k)} - x^{(p)}} \ | \ k \},
        \end{equation}

        When the nearest prototype is updated, all other prototypes should be
        likewise updated. More distant prototypes should move less do nearer
        ones. To quantify how much a prototype should be updated requires not
        only the distance matrix $\mbf{D}_Q$, but also a matrix of the coupling
        strength between prototypes. The \textit{neighborhood matrix} -- $\rm{H}
        \in \mathds{R}^{K \times K}$ -- has elements
        \begin{equation}
            h_{ij} = \exp\left(-\frac{d_{ij}^2}{2\gamma^2}\right),
        \end{equation}
        where $\gamma$ is a user-selected coupling constant. Note that for
        $i=j$, $d_{ij}=1$

        Therefore, for each $k$ the $j$th prototype is updated according to 
        \begin{equation} \label{eq:app:som_update_prototype}
            \mbf{p}^{(k)} \mapsto \mbf{p}^{k} + \alpha h_{c(n), k} \left(\mbf{w}^{(n)} - \mbf{p}^{(c(n))}\right).
        \end{equation}.
        where $\alpha$ is the learning rate, another free parameter.

        When $k=c(n)$, this reduces to the simpler form
        \begin{equation} \label{eq:app:som_BMU_update}
          \mbf{p}^{(k(n))} \rightarrow \mbf{p}^{(k(n))} + \alpha (\mbf{w}^{(n)} - \mbf{p}^{(k(n))}),
        \end{equation}



\label{lastpage}

\end{document}